\documentclass[11pt]{amsart}
\usepackage{amsmath,amssymb,amsthm,mathtools}
\usepackage[margin=1in]{geometry}
\usepackage{booktabs}
\usepackage{longtable}
\usepackage{hyperref}
\usepackage{graphicx}
\usepackage{xcolor}
\usepackage{enumitem}
\usepackage{bm}
\usepackage{tikz}
\usepackage{pgfplots}
\usepackage{listings}
\pgfplotsset{compat=1.18}
\usetikzlibrary{arrows.meta,positioning,calc,decorations.pathmorphing,patterns,shapes.geometric,fit,backgrounds}

\tikzset{
  memristor/.pic={
    \draw[line width=0.8pt] (-0.55,0)--(-0.35,0);
    \draw[line width=0.8pt] (-0.35,-0.16) rectangle (0.35,0.16);
    \fill[black] (0.18,-0.16) rectangle (0.35,0.16);
    \draw[line width=0.8pt] (-0.30,0.10)--(-0.10,0.10)--(-0.10,-0.10)--(0.10,-0.10)--(0.10,0.10)--(0.18,0.10);
    \draw[line width=0.8pt] (0.35,0)--(0.55,0);
  }
}

\newtheorem{theorem}{Theorem}[section]
\newtheorem{proposition}[theorem]{Proposition}

\theoremstyle{definition}

\newtheorem{principle}[theorem]{Design Principle}
\theoremstyle{remark}
\newtheorem{remark}[theorem]{Remark}

\title[Multi-Level Resistive Synapses]{Multi-Level Resistive Synapses for On-Chip Neural Networks: A Physics-Based Design of a Memristive Crossbar Fabric with Quasi-Continuous Conductance States}
\author{David Alejandro Trejo Pizzo}
\thanks{Independent researcher. \texttt{dtrejopizzo@gmail.com}}
\date{October 2025}

\begin{document}
\begin{abstract}
Building on the concept of \emph{resistive communication}, this paper develops a complete physics-based design of an on-chip neural network whose synaptic weights are stored in memristive devices supporting a very large number of sub-levels of resistance. We begin from the ionic transport physics of valence-change and electrochemical-metallization memristors and derive a state-variable model in which the internal filament geometry is a continuous variable. From this model we obtain a programmable conductance with a quasi-continuous spectrum of stable states, and we quantify how many distinguishable sub-levels can be packed into a single device given thermal noise, read disturb, conductance drift, and quantized-conductance effects. We then assemble these devices into a $1$T$1$R crossbar fabric, derive the exact linear algebra of analog vector--matrix multiplication in the presence of wire resistance and sneak paths, and design a differential-pair synapse that maps signed weights onto strictly positive conductances. A multilayer architecture is proposed in which forward inference, backpropagation, and weight update are all executed physically in the analog domain. We derive the in-situ outer-product learning rule, its discretization onto a finite (but very fine) conductance lattice, and the resulting quantization-noise dynamics. We provide an energy and area model, an information-theoretic capacity analysis of the multi-level cell, and a treatment of how resistive communication couples the tiles of a large neuromorphic system. Finally we show that this substrate is especially well suited to executing large language models, and argue quantitatively that it surpasses both binary/few-level ReRAM (by the level-count factor $\log_2 L$ in density and energy) and traditional CMOS (by eliminating weight movement entirely). We detail the material stack (recommending HfO$_2$-based devices), the exact FEOL/BEOL CMOS-integration flow that lets the whole fabric be built in a standard foundry process, a SPICE model and crossbar validation, the complete memristor-based FPGA neuromorphic system, and a ternary BitNet datapath with projected per-token performance several thousand times more energy-efficient than a GPU or an advanced CPU. The result is a self-contained blueprint for a high-density, analog, in-memory neural processor.
\end{abstract}
\maketitle

\setcounter{tocdepth}{1}
\tableofcontents

\section{Introduction}\label{sec:intro}

The end of classical Dennard scaling and the slowing of Moore's law have made the physical separation between memory and computation---the von Neumann bottleneck---the dominant cost of modern data-intensive workloads. Artificial neural networks are the paradigmatic example: the vast majority of their energy is spent moving weights between off-chip memory and arithmetic units rather than performing the multiply--accumulate (MAC) operations themselves \cite{sebastian2020,ielmini2018}.

The memristor, the fourth fundamental two-terminal circuit element \cite{chua1971,strukov2008}, offers a direct escape. Because its conductance is both \emph{programmable} and \emph{persistent}, a memristor can store a synaptic weight and, when a voltage is applied, perform an analog multiplication by Ohm's law, $i = G\,v$. Arranged in a crossbar, a population of such devices performs a full vector--matrix product in a single step, in the place where the data already lives \cite{hu2018,yao2020}.

In previous work \cite{trejo2017} we introduced the idea of \emph{resistive communication}: the observation that the set/reset dynamics of a Mott or valence-change memristor (a ``neuristor'' \cite{pickett2013}) do not merely store a bit but actively transmit information through the formation and rupture of conductive nanofilaments, in a manner analogous to the electrical synapse. That work argued the concept qualitatively. The present paper provides the quantitative, physics-based design that the concept demands, and it focuses on one property that the earlier work only hinted at: the ability of a single device to hold not one bit, but a very large number of \emph{sub-levels} of resistivity.

A device that can occupy, stably and reproducibly, hundreds of distinguishable conductance states is, in effect, an analog synaptic weight. A crossbar of such devices is an analog weight matrix. A stack of such crossbars, wired through resistive-communication channels, is an on-chip neural network.

\subsection{Contributions}

\begin{enumerate}[label=(\roman*)]
\item A continuous state-variable physical model of a multi-level memristor (Section~\ref{sec:physics}), derived from ionic drift--diffusion, from which the quasi-continuous conductance spectrum emerges naturally.
\item A capacity analysis (Section~\ref{sec:capacity}) bounding the number of reliably distinguishable sub-levels per device under thermal noise, drift, quantized conductance, and programming stochasticity.
\item A compact model for circuit simulation (Section~\ref{sec:compact}) and the exact circuit theory of the $1$T$1$R crossbar (Section~\ref{sec:crossbar}), including wire-resistance (IR-drop) corrections and the differential synapse representing signed weights.
\item A complete multilayer architecture (Section~\ref{sec:arch}) performing forward inference, error backpropagation, and weight update entirely in the analog domain, with the in-situ learning rule derived from gradient descent and discretized onto the conductance lattice (Section~\ref{sec:learning}).
\item Mixed-signal peripheral, energy, area, and resistive-communication models tying many tiles into a single fabric (Sections~\ref{sec:periph}--\ref{sec:fabric}).
\item A quantitative case for executing large language models on this fabric (Section~\ref{sec:llm}), establishing its advantage over both binary/few-level ReRAM and conventional CMOS.
\item A material study recommending the optimal stack (Section~\ref{sec:materials}), the exact FEOL/BEOL foundry-integration flow (Section~\ref{sec:cmos}), a SPICE model (Section~\ref{sec:spice}), the complete memristive-FPGA neuromorphic system (Section~\ref{sec:fpga}), and a ternary BitNet datapath with GPU/M4 performance comparisons (Section~\ref{sec:bitnet}).
\item An argument that the design is a \emph{hardware-native} neural-interconnect architecture requiring no high-level program, kernel, or runtime (Section~\ref{sec:native}), and a three-dimensional integration study showing density scaling multiplicatively with tier count (Section~\ref{sec:3d}).
\item A fully in-memory self-attention scheme with current-mode analog softmax (Section~\ref{sec:attention}); a fault-tolerance, yield, and redundancy analysis (Section~\ref{sec:fault}); a memristive physical-unclonable-function for on-chip security (Section~\ref{sec:security}); and a roofline plus an energy--delay optimality theorem positioning the fabric against the fundamental limits (Section~\ref{sec:roofline}).
\end{enumerate}

\subsection{Organization}

Section~\ref{sec:physics} develops the device physics. Section~\ref{sec:capacity} bounds the number of sub-levels. Section~\ref{sec:compact} gives a simulation-ready compact model. Section~\ref{sec:crossbar} treats the crossbar circuit theory. Section~\ref{sec:arch} builds the network architecture, and Section~\ref{sec:learning} the in-situ learning rule. Sections~\ref{sec:periph}--\ref{sec:fabric} cover peripherals, energy/area, and inter-tile resistive communication. Section~\ref{sec:system} addresses system-level mapping and reliability. Sections~\ref{sec:example}--\ref{sec:comparison} give a worked example and a comparison with digital accelerators, and Section~\ref{sec:llm} makes the case for large language models. Section~\ref{sec:conclusion} concludes.

\section{Physics of the Multi-Level Device}\label{sec:physics}

\subsection{From flux--charge to a state variable}

The constitutive relation of the ideal memristor links flux and charge,
\begin{equation}
d\phi = M(q)\,dq,
\end{equation}
so that the charge-controlled memristance $M(q)=d\phi/dq$ depends on the entire history of current that has flowed; equivalently a flux-controlled device has memductance $W(\phi)=dq/d\phi$. No real device is purely charge-controlled: the resistance depends on a physical internal degree of freedom---the geometry of a conductive filament or the position of an oxygen-vacancy front. We therefore adopt the generic \emph{state-variable} formulation,
\begin{align}
i(t) &= G(x,v)\,v(t), \label{eq:ohm}\\
\frac{dx}{dt} &= f(x,v), \label{eq:state}
\end{align}
where $v$ is the terminal voltage, $i$ the current, and $x \in [0,1]$ a dimensionless internal state encoding, e.g., the normalized length of the conductive filament.

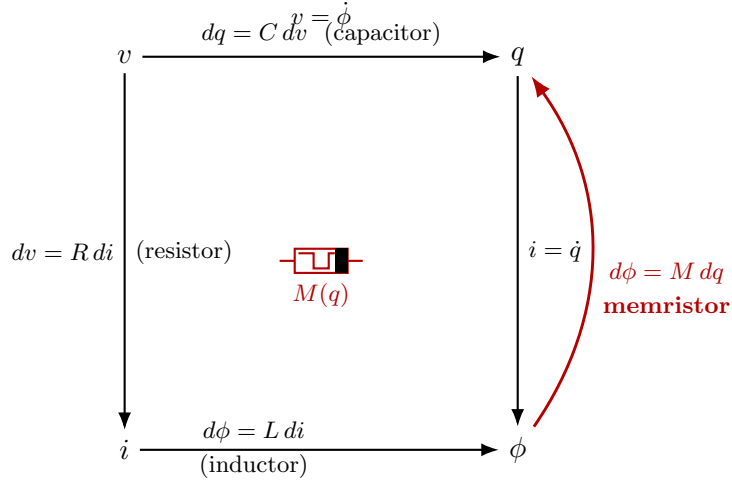
\begin{figure}[h!]
\centering
\begin{tikzpicture}[>=Latex,scale=1.0]
  \node (v) at (-2.6,2.6) {$v$};
  \node (q) at (2.6,2.6) {$q$};
  \node (i) at (-2.6,-2.6) {$i$};
  \node (phi) at (2.6,-2.6) {$\phi$};
  \draw[->,thick] (v) -- node[above]{\footnotesize $dq=C\,dv$ \ (capacitor)} (q);
  \draw[->,thick] (i) -- node[left,align=center]{\footnotesize $d\phi=L\,di$\\\footnotesize (inductor)} (phi);
  \draw[->,thick] (v) -- node[left]{\footnotesize $dv=R\,di$} node[right]{\footnotesize (resistor)} (i);
  \draw[->,thick] (q) -- node[right,align=center]{\footnotesize $i=\dot q$} (phi);
  \draw[->,line width=1.1pt,red!70!black]
       (phi) to[bend right=35] node[below right,align=center]{\footnotesize $d\phi=M\,dq$\\\footnotesize \textbf{memristor}} (q);
  \node[align=center] at (0,3.15) {\footnotesize $v=\dot\phi$};
  \draw[red!70!black,thick] (0,-0.1) pic[scale=1.0]{memristor};
  \node[red!70!black] at (0,-0.55){\footnotesize $M(q)$};
\end{tikzpicture}
\caption{The four fundamental two-terminal elements as relations among the pairs $(v,i,q,\phi)$.}
\label{fig:elements}
\end{figure}

Equations~\eqref{eq:ohm}--\eqref{eq:state} are the foundation of everything that follows: \eqref{eq:ohm} is the \emph{read} (multiplication) path used during inference; \eqref{eq:state} is the \emph{write} (programming) path used during learning. The central design problem is to make $x$---and hence $G$---occupy a dense set of stable values.

\begin{principle}\label{prin:dense}
A useful analog synapse requires that the map $x \mapsto G(x)$ be continuous and that the dynamics \eqref{eq:state} admit a dense set of \emph{metastable} fixed points under the zero-bias (retention) condition $f(x,0)\approx 0$. The number of such metastable states is the number of storable sub-levels.
\end{principle}

\subsection{Linear and nonlinear drift models}

In the Strukov $\mathrm{TiO_2}$ picture \cite{strukov2008} the device is a series combination of a doped (low-resistance) region of width $w=xD$ and an undoped (high-resistance) region of width $(1-x)D$, where $D$ is the film thickness:
\begin{equation}
R(x) = R_\mathrm{on}\,x + R_\mathrm{off}\,(1-x).
\label{eq:Rlin}
\end{equation}
The dopant front drifts under the local field with ionic mobility $\mu_v$,
\begin{equation}
\frac{dx}{dt} = \frac{\mu_v R_\mathrm{on}}{D^2}\, i(t)
              = \frac{\mu_v R_\mathrm{on}}{D^2}\, \frac{v(t)}{R(x)}.
\label{eq:drift}
\end{equation}
The factor $1/D^2$ is the origin of the nanoscale signature of the memristor: shrinking $D$ from microns to nanometers amplifies the state's sensitivity to charge by twelve orders of magnitude, which is precisely why memristive memory is a nanoscale phenomenon.

Equation~\eqref{eq:drift} alone drives the state to its rails, so a \emph{window function} $g(x)$ is introduced to enforce $x\in[0,1]$ and model the vanishing of ionic drift near the boundaries:
\begin{equation}
\frac{dx}{dt} = \eta\,\frac{\mu_v R_\mathrm{on}}{D^2}\,\frac{v}{R(x)}\, g(x),
\qquad g(x)=1-(2x-1)^{2p},
\label{eq:window}
\end{equation}
with polarity $\eta=\pm1$ and window sharpness $p$. The window captures the fact that programming becomes progressively harder near either extreme, which directly shapes the spacing of the achievable sub-levels.

\subsection{Drift--diffusion of mobile ions: the Nernst--Planck picture}

The phenomenological window model can be grounded in the microscopic transport of the mobile species (oxygen vacancies, or metal cations in CBRAM) \cite{waser2007,yang2008}. Let $n(\mathbf r,t)$ be the vacancy concentration. Conservation of the ionic species obeys the continuity equation $\partial_t n + \nabla\!\cdot\mathbf J_\mathrm{ion}=0$, with the Nernst--Planck flux combining drift and Fickian diffusion,
\begin{equation}
\mathbf J_\mathrm{ion}
= \underbrace{\mu_v\, n\,\mathbf E}_{\text{drift}}
- \underbrace{D_v\,\nabla n}_{\text{diffusion}},
\qquad D_v = \frac{\mu_v k_B T}{q_i},
\label{eq:nernstplanck}
\end{equation}
the diffusivity and mobility being tied by the Einstein relation. The internal state $x$ of the compact model is the spatial average of $n$ over the active region $\Omega$,
\begin{equation}
x(t) = \frac{1}{n_0\,\Omega}\int_\Omega n(\mathbf r,t)\,d^3r,
\end{equation}
so integrating \eqref{eq:nernstplanck} over $\Omega$ and applying the divergence theorem reproduces a first-order law $\dot x = f(x,v)$: the drift term gives the field-driven set/reset and the diffusion term the spontaneous relaxation (retention loss). The compact \eqref{eq:window} is thus the volume-averaged Nernst--Planck equation, and the window $g(x)$ encodes the geometric confinement of $\Omega$.

\subsection{Ionic transport: threshold and the exponential write}

The linear model hides the strong nonlinearity that makes multi-level storage possible. Ion hopping over an energy barrier $U_a$ in a field $E$ follows a Mott--Gurney / Butler--Volmer form; the drift velocity of the vacancy front is
\begin{equation}
\frac{dx}{dt} = \nu_0\, e^{-U_a/k_BT}\,
\sinh\!\left(\frac{q_i a E}{2 k_B T}\right),
\label{eq:hopping}
\end{equation}
with attempt frequency $\nu_0$, hopping distance $a$, ionic charge $q_i$. Writing $E=v/D$ and grouping constants,
\begin{equation}
\frac{dx}{dt} = \kappa\, e^{-U_a/k_BT}\,
\sinh(v/V_0), \qquad V_0 = \frac{2k_BTD}{q_i a}.
\label{eq:sinh}
\end{equation}

\begin{proposition}[Non-destructive read]\label{prop:read}
For a read pulse with $|v|\ll V_0$ the relative state perturbation per read is
\[
\frac{|\Delta x|}{\Delta x_\mathrm{set}} \;\approx\; \frac{\sinh(v/V_0)}{\sinh(v_p/V_0)} \;\ll\; 1,
\]
where $v_p$ is the program voltage. Hence the stored sub-level is read non-destructively, and the number of reads before a refresh is required scales as $\sinh(v_p/V_0)/\sinh(v_r/V_0)$.
\end{proposition}

The steep $\sinh$ thus does double duty: below threshold the state is effectively frozen (retention and non-destructive read), and above it the state moves exponentially fast (fast write). This is the single most important physical fact enabling a dense analog memory.

\subsection{Joule self-heating and thermal runaway of reset}

The barrier $U_a$ in \eqref{eq:sinh} is for hopping at the lattice temperature, but the filament heats under current. A lumped energy balance gives
\begin{equation}
C_\mathrm{th}\frac{dT}{dt}
= \underbrace{G(x)\,v^2}_{\text{Joule input}}
- \underbrace{\frac{T-T_0}{R_\mathrm{th}}}_{\text{conduction to bath}},
\label{eq:thermal}
\end{equation}
with thermal capacitance $C_\mathrm{th}$ and resistance $R_\mathrm{th}$. In quasi-static operation $T = T_0 + R_\mathrm{th}G v^2$, and substituting into the Arrhenius factor couples electrical and thermal states:
\begin{equation}
\dot x \propto \exp\!\Bigl(-\frac{U_a}{k_B(T_0+R_\mathrm{th}G v^2)}\Bigr)\sinh(v/V_0).
\end{equation}
This positive feedback---more current, more heat, faster ion motion---makes the \emph{reset} of unipolar devices abrupt and sets the compliance current the access transistor must enforce. For multi-level operation we deliberately remain in the bipolar, field-driven regime where the $\sinh$ term dominates and thermal feedback is weak, because abrupt thermal reset destroys intermediate levels.

\subsection{Filamentary conductance and quantization}

For electrochemical-metallization (CBRAM) and many oxide devices, conduction is filamentary: a metallic bridge of cross-section $A_f(x)$ shorts the electrodes, with
\begin{equation}
G(x) = \sigma_f \frac{A_f(x)}{\ell_f} + G_\mathrm{leak}.
\end{equation}
When the filament narrows to atomic dimensions, conduction is ballistic and the conductance is quantized in units of the conductance quantum \cite{siemon2016},
\begin{equation}
G_0 = \frac{2e^2}{h} \approx 77.5\ \mu\mathrm{S}\quad(\,\approx (12.9\ \mathrm{k}\Omega)^{-1}\,),
\end{equation}
so near the bottom of the range $G \simeq n\,G_0$ with integer $n$. This is both a blessing---naturally reproducible low-conductance sub-levels at $1G_0, 2G_0,\dots$---and a curse---a hard granularity at the very lowest states. The usable analog regime therefore lives \emph{above} the few-atom quantized regime, where many parallel channels average into a quasi-continuous $G(x)$:
\begin{equation}
G \in [\,G_\mathrm{min},\,G_\mathrm{max}\,],\qquad G_\mathrm{min} \gtrsim 10\,G_0.
\end{equation}

\subsection{Electronic conduction mechanisms}

Between the metallic LRS and the insulating HRS the current is carried by several
competing electronic transport channels, each with a distinct $i$--$v$ signature
that the compact model of Section~\ref{sec:compact} must reproduce. In the HRS the
residual tunneling gap $d_g$ dominates and conduction is \emph{Simmons/Fowler--Nordheim}
direct tunneling,
\begin{equation}
i_\mathrm{tun} \propto \frac{v}{d_g}\exp\!\Bigl(-\tfrac{4\pi d_g}{h}\sqrt{2m^{*}\Phi_B}\Bigr),
\end{equation}
exponentially sensitive to the barrier width $d_g$ and height $\Phi_B$---which is
exactly why a sub-nanometre change in the filament gap produces a large, and
finely controllable, conductance change. At higher fields trap-assisted
\emph{Poole--Frenkel} emission contributes,
\begin{equation}
i_\mathrm{PF}\propto v\,\exp\!\Bigl(\frac{-q(\Phi_T-\sqrt{qv/\pi\varepsilon d})}{k_BT}\Bigr),
\end{equation}
with trap depth $\Phi_T$ and permittivity $\varepsilon$, while bulk-limited
\emph{space-charge-limited conduction} gives the Mott--Gurney law
$i_\mathrm{SCLC}=\tfrac98\varepsilon\mu\,v^2/d^3$. The superposition of these
channels, each thermally activated, is what makes $G(x)$ a smooth, monotone, and
densely tunable function of the internal state---the electronic underpinning of
Design Principle~\ref{prin:dense}.

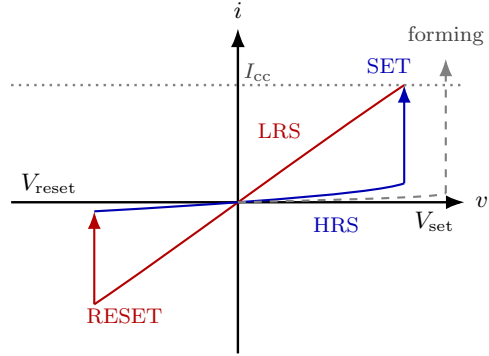
\begin{figure}[h!]
\centering
\begin{tikzpicture}[>=Latex,line width=0.9pt]
\begin{scope}
  \draw[->] (-3.0,0)--(3.0,0) node[right,font=\footnotesize]{$v$};
  \draw[->] (0,-2.0)--(0,2.3) node[above,font=\footnotesize]{$i$};
  \node[font=\footnotesize] at (2.6,-0.25){$V_\mathrm{set}$};
  \node[font=\footnotesize] at (-2.5,0.25){$V_\mathrm{reset}$};
  \draw[blue!70!black,thick] (0,0) .. controls (1.4,0.10) and (2.0,0.18) .. (2.2,0.25);
  \draw[blue!70!black,thick,->] (2.2,0.25)--(2.2,1.55);
  \node[font=\scriptsize,blue!70!black] at (2,1.8){SET};
  \draw[red!70!black,thick] (2.2,1.55) .. controls (1.0,0.75) and (-1.0,-0.75) .. (-1.9,-1.35);
  \draw[red!70!black,thick,->] (-1.9,-1.35)--(-1.9,-0.12);
  \node[font=\scriptsize,red!70!black] at (-1.5,-1.5){RESET};
  \draw[blue!70!black,thick] (-1.9,-0.12) .. controls (-1.0,-0.07) and (-0.4,-0.03) .. (0,0);
  \draw[gray,thick,dashed] (0,0) .. controls (2.2,0.05) and (2.7,0.08) .. (2.75,0.12);
  \draw[gray,thick,dashed,->] (2.75,0.12)--(2.75,1.9);
  \node[font=\scriptsize,gray!50!black] at (2.75,2.2){forming};
  \draw[gray,dotted] (-3,1.55)--(3,1.55);
  \node[font=\scriptsize,gray!50!black,anchor=east] at (0.6,1.72){$I_\mathrm{cc}$};
  \node[font=\scriptsize,blue!70!black] at (1.3,-0.30){HRS};
  \node[font=\scriptsize,red!70!black] at (0.55,0.95){LRS};
\end{scope}
\end{tikzpicture}
\caption{Bipolar resistive-switching $i$--$v$ loop. From the pristine cell a
\emph{forming} sweep (dashed) nucleates the first filament. Thereafter a positive
\emph{SET} at $V_\mathrm{set}$ switches HRS$\to$LRS (current capped at the
compliance $I_\mathrm{cc}$ supplied by the $1$T$1$R transistor), and a negative
\emph{RESET} at $V_\mathrm{reset}$ ruptures the filament LRS$\to$HRS. Stopping
the SET/RESET partway places the cell at an intermediate conductance---one analog
sub-level.}
\label{fig:switching}
\end{figure}

\subsection{Materials of the memristor and the optimal stack}\label{sec:materials}

The electrical behaviour derived above is set by the material stack. A
valence-change memristor is a metal--insulator--metal (MIM) sandwich whose
switching layer is a sub-stoichiometric transition-metal oxide; the mobile species
are oxygen vacancies $V_O^{\bullet\bullet}$ created and annihilated at an
oxygen-exchange interface. The relevant material parameters map directly onto the
model: the oxide bandgap $E_g$ and electron affinity set the barrier height
$\Phi_B$; the vacancy formation enthalpy and migration barrier set $U_a$ in
\eqref{eq:sinh}; the dielectric constant $\varepsilon_r$ sets the field for a given
voltage; and the electrode work functions set the Schottky asymmetry that makes
switching bipolar.

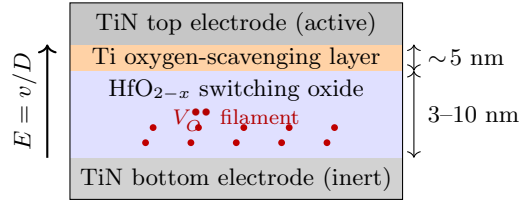
\begin{figure}[h!]
\centering
\begin{tikzpicture}[font=\footnotesize]
  \def\w{4.4}
  \fill[gray!35] (0,0)   rectangle (\w,0.55);  \node at (\w/2,0.275){TiN bottom electrode (inert)};
  \fill[blue!12] (0,0.55) rectangle (\w,1.7);   \node at (\w/2,1.45){HfO$_{2-x}$ switching oxide};
  \foreach \x in {1.0,1.6,2.2,2.8,3.4}{ \fill[red!75!black] (\x,0.75) circle (0.045);
     \fill[red!75!black] (\x+0.1,0.95) circle (0.045);}
  \node[red!70!black] at (\w/2,1.05){\scriptsize $V_O^{\bullet\bullet}$ filament};
  \fill[orange!35] (0,1.7) rectangle (\w,2.05); \node at (\w/2,1.875){Ti oxygen-scavenging layer};
  \fill[gray!45] (0,2.05) rectangle (\w,2.6);  \node at (\w/2,2.325){TiN top electrode (active)};
  \draw (0,0) rectangle (\w,2.6);
  \draw[<->] (\w+0.15,0.55)--(\w+0.15,1.7); \node[right] at (\w+0.2,1.12){$3$--$10$ nm};
  \draw[<->] (\w+0.15,1.7)--(\w+0.15,2.05);  \node[right] at (\w+0.2,1.875){$\sim\!5$ nm};
  \draw[->,thick] (-0.3,0.55)--(-0.3,2.05) node[midway,left,rotate=90,anchor=south]{$E=v/D$};
\end{tikzpicture}
\caption{A reactive Ti oxygen-scavenging layer between the active TiN electrode and the HfO$_{2-x}$ switching oxide creates and stores oxygen, leaving a vacancy-rich region that nucleates the conductive filament; the inert bottom TiN electrode anchors it. All four materials are already qualified in standard logic BEOL.}
\label{fig:stack}
\end{figure}

\begin{table}[h!]
\centering
\caption{Candidate switching oxides for an analog multi-level synapse.}\label{tab:materials}
\vspace{0.15cm}
\begin{tabular}{@{}lccccp{2.6cm}@{}}
\toprule
\textbf{Oxide} & \textbf{$E_g$} & \textbf{Endurance} & \textbf{Levels} & \textbf{CMOS} & \textbf{Note} \\
\midrule
HfO$_2$   & $5.7$ eV & $10^{6}$--$10^{9}$ & high & \textbf{native} & high-$k$ since 45\,nm HKMG \\
Ta$_2$O$_5$/TaO$_x$ & $4.4$ eV & $10^{10}$--$10^{12}$ & high & good & best endurance \\
TiO$_2$   & $3.2$ eV & $10^{6}$ & med & ok & the original \cite{strukov2008} \\
Al$_2$O$_3$ & $6.4$ eV & $10^{8}$ & low & good & barrier/series layer \\
SiO$_x$   & $8.9$ eV & $10^{7}$ & med & native & fully Si-compatible \\
\bottomrule
\end{tabular}
\end{table}

\begin{principle}[Best material choice]\label{prin:material}
For foundry integration the unambiguous primary choice is \emph{hafnium oxide}
(HfO$_{2-x}$) with TiN electrodes and a thin Ti oxygen-scavenging layer
(Fig.~\ref{fig:stack}). HfO$_2$ is already the high-$k$ gate dielectric in every
high-$k$/metal-gate (HKMG) logic node since $45$~nm, so its deposition (ALD),
etch, and reliability are fully qualified at every major foundry; TiN is the
standard barrier/electrode metal. Where maximum endurance is needed (frequent
in-situ training) a \emph{bilayer} HfO$_x$/TaO$_x$ stack is preferred, adding the
$10^{12}$-cycle endurance of tantalum oxide while retaining a Hf-based,
CMOS-clean interface. We therefore specify HfO$_{2-x}$ (optionally
HfO$_x$/TaO$_x$ bilayer) for the remainder of the paper.
\end{principle}

The atomic-layer-deposited HfO$_2$ thickness $D$ enters the model through the
$1/D^2$ sensitivity of \eqref{eq:drift} and the threshold $V_0\propto D$ of
\eqref{eq:sinh}; a $5$~nm film gives a $\sim\!1$~V programming threshold and a
sub-volt read, matching the supply rails of a standard logic process.

\subsection{The conductance spectrum}

Combining \eqref{eq:Rlin} with a filamentary correction, the steady-state conductance is well approximated by a geometric (log-linear) law,
\begin{equation}
G(x) = G_\mathrm{min}\left(\frac{G_\mathrm{max}}{G_\mathrm{min}}\right)^{x},
\label{eq:Gexp}
\end{equation}
because filament cross-section and tunneling gap both depend exponentially on the displaced ionic charge. The \emph{dynamic range} is
\begin{equation}
\rho \equiv \frac{G_\mathrm{max}}{G_\mathrm{min}} = \frac{R_\mathrm{off}}{R_\mathrm{on}} \in [10,\,10^3].
\end{equation}
Equation~\eqref{eq:Gexp} says equal increments of stored charge produce equal \emph{ratios} of conductance---ideal for representing weights on a logarithmic scale; for a linear weight scale the controller pre-distorts the pulse train (Section~\ref{sec:learning}).

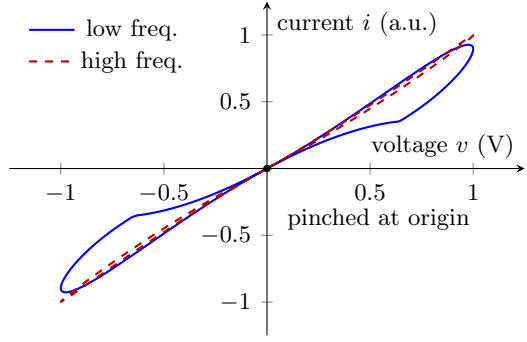
\begin{figure}[h!]
\centering
\begin{tikzpicture}
\begin{axis}[
  width=8.4cm,height=6.0cm,
  xlabel={voltage $v$ (V)},ylabel={current $i$ (a.u.)},
  axis lines=middle,
  xmin=-1.25,xmax=1.25,ymin=-1.25,ymax=1.25,
  xtick={-1,-0.5,0.5,1},ytick={-1,-0.5,0.5,1},
  tick label style={font=\footnotesize},
  label style={font=\footnotesize},
  samples=200,domain=0:360,
  legend style={font=\footnotesize,at={(0.02,0.98)},anchor=north west,draw=none,fill=none},
]
\addplot[blue,thick,smooth] ({sin(x)},{ sin(x)*(0.55+0.45*abs(sin(x-40)))});
\addplot[red!70!black,thick,dashed,smooth] ({sin(x)},{ sin(x)*(0.85+0.15*abs(sin(x-12)))});
\legend{low freq.,high freq.}
\node[circle,fill,inner sep=1pt] at (axis cs:0,0){};
\node[anchor=west,font=\footnotesize] at (axis cs:0.05,-0.38){pinched at origin};
\end{axis}
\end{tikzpicture}
\caption{Pinched-hysteresis $I$--$V$ signature of a memristor: the loop passes through the origin (zero current at zero voltage). The lobe area shrinks as the drive frequency rises (dashed) because the state variable $x$ has less time to respond---the fingerprint of $x$ lagging the drive in \eqref{eq:state}. Compact-model parameters of Section~\ref{sec:compact} are extracted by matching this lobe area versus frequency.}
\label{fig:iv}
\end{figure}

\section{How Many Sub-Levels? A Capacity Analysis}\label{sec:capacity}

A sub-level is \emph{usable} only if it can be (i) written reproducibly, (ii) retained, and (iii) read back distinguishably from its neighbors. We bound the number $L$ of usable levels.

\subsection{Read noise floor}

During a read at bias $V_r$, the current $I=GV_r$ is corrupted by thermal (Johnson--Nyquist) noise; over a read bandwidth $\Delta f$ the thermal current-noise variance is $\sigma_{I,\mathrm{th}}^2 = 4 k_B T\, G\, \Delta f$. Referred to conductance, $\sigma_G=\sigma_I/V_r$,
\begin{equation}
\sigma_{G,\mathrm{th}} = \frac{1}{V_r}\sqrt{4 k_B T\, G\, \Delta f}.
\label{eq:sigmaG}
\end{equation}
To separate two adjacent levels with spacing $\Delta G$ at a target bit-error rate we require a margin of $m$ standard deviations (e.g.\ $m=6$ for $\mathrm{BER}\!\sim\!10^{-9}$): $\Delta G \ge m\,\sigma_{G,\mathrm{th}}$.

\subsection{Spectral view of the read noise}

The full read-noise PSD sums thermal, shot, and flicker contributions,
\begin{equation}
S_I(f) = \underbrace{4k_BT G}_{\text{thermal}} + \underbrace{2qI}_{\text{shot}} + \underbrace{\frac{K_f I^{2}}{f^{\alpha}}}_{\text{flicker }(1/f)},\quad \alpha\approx1,
\end{equation}
giving integrated noise over $[f_1,f_2]$,
\begin{equation}
\sigma_I^2 = (4k_BTG + 2qI)(f_2-f_1) + K_f I^2 \ln\frac{f_2}{f_1}.
\end{equation}
At small read currents the $\propto\!I^2$ flicker term dominates at low frequency, motivating \emph{correlated double sampling} in the readout, which subtracts the slow $1/f$ component and recovers the white-limited resolution assumed below.

\subsection{Number of distinguishable levels}

Spacing the levels geometrically (constant $\Delta G/G$, matching \eqref{eq:Gexp}) and integrating the resolution constraint across the dynamic range gives a channel-capacity-like bound.

\begin{theorem}[Sub-level count]\label{thm:levels}
Under a thermal read-noise floor and an $m$-sigma separation margin, the number of distinguishable conductance levels is
\begin{equation}
L_\mathrm{th} = 1 + \frac{V_r}{m\sqrt{4k_BT\,\Delta f}}\int_{G_\mathrm{min}}^{G_\mathrm{max}}\frac{dG}{\sqrt{G}}
= 1 + \frac{2V_r\bigl(\sqrt{G_\mathrm{max}}-\sqrt{G_\mathrm{min}}\bigr)}{m\sqrt{4k_BT\,\Delta f}}.
\label{eq:Lth}
\end{equation}
Equivalently the per-read information capacity is $C=\log_2 L = \tfrac12\log_2(1+S/N)$, with $S/N$ the squared ratio of full-scale to RMS read current.
\end{theorem}

\begin{remark}
As a representative point, with $V_r=0.2$~V, $G_\mathrm{max}=200~\mu$S, $G_\mathrm{min}=2~\mu$S, $T=300$~K, $\Delta f=50$~MHz (matching a $\tau_r\!\approx\!10$~ns read), and $m=6$, \eqref{eq:Lth} yields $L_\mathrm{th}\approx9\times10^{2}$ thermally-distinguishable levels---on the order of $10$ effective bits per cell. Device drift and programming stochasticity then reduce the \emph{usable} count to several hundred ($\sim\!8$ bits), consistent with experimentally reported multilevel ReRAM/PCM cells \cite{ambrogio2018,boybat2018}.
\end{remark}

\begin{table}[h!]
\centering
\caption{Representative multi-level cell budget (bipolar oxide device).}\label{tab:budget}
\vspace{0.15cm}
\begin{tabular}{@{}llp{6.2cm}@{}}
\toprule
\textbf{Quantity} & \textbf{Value} & \textbf{Role} \\
\midrule
$G_\mathrm{max}/G_\mathrm{min}$ & $100$ & Dynamic range $\rho$; sets weight range \eqref{eq:Gexp} \\
$V_r$ & $0.2$ V & Read bias; below threshold $V_0$ (Prop.~\ref{prop:read}) \\
$V_0$ & $\sim\!1$ V & Programming threshold scale \eqref{eq:sinh} \\
$\Delta f$ & $50$ MHz & Read bandwidth; sets thermal floor \eqref{eq:sigmaG} \\
$m$ & $6$ & Separation margin ($\mathrm{BER}\!\sim\!10^{-9}$) \\
$L_\mathrm{th}$ & $\sim\!930$ & Thermal-limit levels (Thm.~\ref{thm:levels}) \\
$L_\mathrm{usable}$ & $\sim\!256$ & After drift + programming noise \\
$\log_2 L$ & $\sim\!8$ bits & Usable per-cell resolution \\
\bottomrule
\end{tabular}
\end{table}

\subsection{Drift, retention, and the density--robustness trade}

Real devices relax: after programming the filament reconfigures and $G$ drifts, often as a power law $G(t)=G(t_0)(t/t_0)^{-\nu_d}$ with a small exponent $\nu_d$. Levels spaced more finely than the drift spread $\Delta G_\mathrm{drift}(t)$ merge before the next refresh, so the effective level count is
\begin{equation}
L_\mathrm{eff} = \min\!\Bigl(L_\mathrm{th},\;
\frac{\ln \rho}{\ln\bigl(1+\nu_d \ln(t_\mathrm{ret}/t_0)\bigr)}\Bigr),
\end{equation}
formalizing the trade between density (many fine levels) and retention (coarse, robust levels). For neural-network weights this trade is benign: inference tolerates weight perturbations, and periodic in-situ training (Section~\ref{sec:learning}) continually refreshes the array.

\subsection{Programming stochasticity}

Each program pulse moves $x$ by a stochastic increment because ion hopping is Poisson-like. From \eqref{eq:sinh}, a pulse of width $\tau$ has
\begin{equation}
\mu_{\Delta x} = \kappa e^{-U_a/k_BT}\sinh(v_p/V_0)\,\tau,\qquad
\sigma_{\Delta x}^2 = \mu_{\Delta x}/N_\mathrm{ion},
\end{equation}
where $N_\mathrm{ion}$ is the number of mobile ions; larger filaments write more deterministically. \emph{Program-and-verify} suppresses this: after each pulse the controller reads $G$ and adjusts the next pulse, converging the realized level with residual variance set by read noise \eqref{eq:sigmaG} rather than write stochasticity. This is how hundreds of analog levels are achieved in practice \cite{alibart2013}.

\subsection{Mutual-information capacity of one cell}

The level count \eqref{eq:Lth} is a hard-decision bound. The true storage capacity is the mutual information between the programmed weight $w$ and the read-back $\hat w=w+\zeta$, where $\zeta$ aggregates read noise, drift, and quantization with variance $\sigma_\zeta^2$:
\begin{equation}
C_\mathrm{cell}=\max_{p(w)} I(w;\hat w) = \tfrac12\log_2\!\Bigl(1+\frac{\sigma_w^2}{\sigma_\zeta^2}\Bigr)\ \text{bits},
\end{equation}
maximized by a Gaussian source, with $L=2^{C_\mathrm{cell}}$. Every halving of $\sigma_\zeta$ buys one extra bit (a doubling of usable sub-levels): dense multi-level storage \emph{is} the program of driving $\sigma_\zeta$ down through differential sensing, correlated double sampling, and program-and-verify.

\section{A Compact Model for Simulation}\label{sec:compact}

For circuit-level (SPICE) design the distributed physics is collapsed into a two-equation compact model. The read path interpolates between tunneling-limited HRS and ohmic LRS,
\begin{equation}
i = (1-x)\,\gamma\sinh(v/v_t) + x\,G_\mathrm{on}v,
\end{equation}
where the first term is Simmons-barrier tunneling through the residual gap ($\gamma,v_t$ fit parameters) and the second is metallic filament conduction. The state evolves by the thresholded drift
\begin{equation}
\frac{dx}{dt} = \mu_v\,\mathrm{sgn}(v)\,(|v|-v_\mathrm{th})_+^{\,\alpha}\,g(x),
\qquad (\cdot)_+=\max(\cdot,0),
\label{eq:compactstate}
\end{equation}
with programming threshold $v_\mathrm{th}$ and field exponent $\alpha$. Table-free, differentiable, and continuous, this model is what the architecture simulator of Section~\ref{sec:arch} integrates per device; its parameters are extracted from the pinched loop of Fig.~\ref{fig:iv} by matching loop area versus frequency.

\section{The Crossbar Fabric}\label{sec:crossbar}

\subsection{Ideal analog vector--matrix multiplication}

Place memristors at every junction of an $N\times M$ grid, device $(i,j)$ having conductance $G_{ij}$. Drive row $i$ with voltage $v_i$ and hold every column at virtual ground (a transimpedance amplifier). By Kirchhoff's current law the current summed onto column $j$ is
\begin{equation}
I_j = \sum_{i=1}^{N} G_{ij}\, v_i,\qquad \mathbf{I} = \mathbf{G}^{\!\top}\mathbf{v}.
\label{eq:vmm}
\end{equation}

\begin{theorem}[Analog VMM primitive]\label{thm:vmm}
The crossbar of \eqref{eq:vmm} computes a complete vector--matrix product in $O(1)$ time using $O(NM)$ devices, dissipating energy only in the devices themselves. A neural layer's pre-activation $\mathbf z=\mathbf W^{\!\top}\mathbf x$ is realized by encoding $\mathbf x$ as voltages and reading $\mathbf z$ as column currents.
\end{theorem}

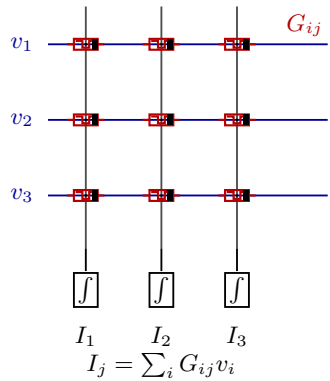
\begin{figure}[h!]
\centering
\begin{tikzpicture}[>=Latex,line width=0.7pt,scale=1.0]
  \def\nx{3} \def\ny{3}
  \foreach \r/\lab in {0/1,1/2,2/3}{
    \pgfmathsetmacro\yy{2-\r}
    \draw[blue!60!black] (-0.4,\yy)--(3.3,\yy);
    \node[left,blue!60!black,font=\footnotesize] at (-0.45,\yy){$v_{\lab}$};
  }
  \foreach \c in {0,1,2}{
    \draw[black!70] (\c+0.1,2.5)--(\c+0.1,-0.7);
  }
  \foreach \r in {0,1,2}{ \foreach \c in {0,1,2}{
    \pgfmathsetmacro\yy{2-\r}
    \draw[red!70!black] (\c+0.1,\yy) pic[scale=0.45]{memristor};
    \draw[red!70!black] (\c+0.1,\yy)--(\c+0.1,\yy);
  }}
  \foreach \r in {0,1,2}{ \foreach \c in {0,1,2}{
    \pgfmathsetmacro\yy{2-\r}
    \fill (\c+0.1,\yy) circle (0.5pt);
  }}
  \foreach \c/\lab in {0/1,1/2,2/3}{
    \draw (\c+0.1,-0.7)--(\c+0.1,-1.0);
    \node[draw,fill=white,inner sep=1.4pt,font=\footnotesize] (op\c) at (\c+0.1,-1.25){$\int$};
    \node[below,font=\footnotesize] at (\c+0.1,-1.6){$I_{\lab}$};
  }
  \node[font=\footnotesize] at (1.1,-2.25){$I_j=\sum_i G_{ij}v_i$};
  \node[red!70!black,font=\footnotesize] at (3.0,2.25){$G_{ij}$};
\end{tikzpicture}
\caption{Analog vector--matrix multiplication in a memristive crossbar. Input voltages $v_i$ drive the rows; each junction memristor $G_{ij}$ injects a current $G_{ij}v_i$ onto its column; Kirchhoff's law sums them, and a transimpedance integrator reads $I_j=\sum_i G_{ij}v_i$. The whole product $\mathbf I=\mathbf G^{\!\top}\mathbf v$ is computed in one step (Theorem~\ref{thm:vmm}).}
\label{fig:crossbar}
\end{figure}

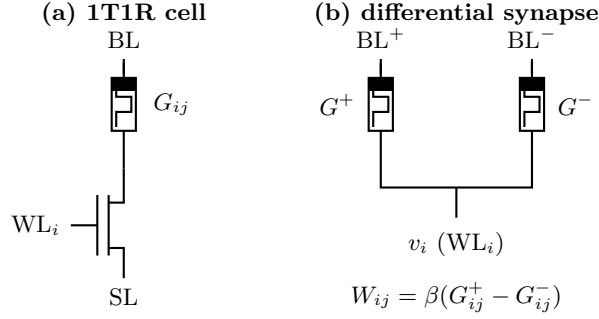
\begin{figure}[h!]
\centering
\begin{tikzpicture}[>=Latex,line width=0.8pt]
  \begin{scope}[shift={(0,0)}]
    \node[font=\footnotesize\bfseries] at (0.0,2.6){(a) 1T1R cell};
    \draw (0,2.0)--(0,1.8); \node[above,font=\footnotesize] at (0,2.0){BL};
    \draw (0,1.4) pic[rotate=90]{memristor};
    \node[right,font=\footnotesize] at (0.25,1.4){$G_{ij}$};
    \draw (0,0.85)--(0,0.55);
    \draw (0,0.55)--(0,0.2);                       
    \draw (-0.35,0.2)--(-0.35,-0.6);               
    \draw (-0.7,-0.2)--(-0.35,-0.2);               
    \node[left,font=\footnotesize] at (-0.7,-0.2){WL$_i$};
    \draw (-0.2,0.2)--(-0.2,-0.6);                 
    \draw (-0.2,0.1)--(0,0.1)--(0,0.2);            
    \draw (-0.2,-0.5)--(0,-0.5)--(0,-0.9);         
    \node[below,font=\footnotesize] at (0,-0.9){SL};
  \end{scope}
  \begin{scope}[shift={(4.2,0)}]
    \node[font=\footnotesize\bfseries] at (0.2,2.6){(b) differential synapse};
    \draw (-0.8,2.0)--(-0.8,1.8); \node[above,font=\footnotesize] at (-0.8,2.0){$\mathrm{BL}^+$};
    \draw (1.2,2.0)--(1.2,1.8);  \node[above,font=\footnotesize] at (1.2,2.0){$\mathrm{BL}^-$};
    \draw (-0.8,1.4) pic[rotate=90]{memristor};
    \draw (1.2,1.4)  pic[rotate=90]{memristor};
    \node[left,font=\footnotesize]  at (-1.0,1.4){$G^+$};
    \node[right,font=\footnotesize] at (1.4,1.4){$G^-$};
    \draw (-0.8,0.85)--(-0.8,0.3)--(0.2,0.3);
    \draw (1.2,0.85)--(1.2,0.3)--(0.2,0.3);
    \draw (0.2,0.3)--(0.2,-0.1);
    \node[font=\footnotesize] at (0.2,-0.45){$v_i$ (WL$_i$)};
    \node[align=center,font=\footnotesize] at (0.2,-1.15){$W_{ij}=\beta(G^+_{ij}-G^-_{ij})$};
  \end{scope}
\end{tikzpicture}
\caption{(a) The $1$T$1$R synapse: an access transistor gates the memristor $G_{ij}$, suppressing sneak paths and supplying the set compliance current $I_\mathrm{cc}$ that sets $G_\mathrm{max}$. (b) The differential synapse encodes a signed weight as $W_{ij}=\beta(G^+_{ij}-G^-_{ij})$ on two bit lines; the two devices age together, cancelling common-mode drift.}
\label{fig:synapse}
\end{figure}

\subsection{Signed weights: the differential synapse}

Conductances are positive but weights are signed. Represent each weight by a \emph{differential pair} on two adjacent columns,
\begin{equation}
W_{ij} = \beta\,(G_{ij}^{+} - G_{ij}^{-}),
\label{eq:diff}
\end{equation}
with $\beta$ a fixed transconductance scale; the column currents subtract in the analog domain, $z_j=\beta\sum_i(G_{ij}^{+}-G_{ij}^{-})v_i$. This doubles device count but yields (i) a full signed range, (ii) first-order cancellation of common-mode drift and temperature, since $G^+$ and $G^-$ age together, and (iii) a robust zero at $G^+=G^-$ anywhere in range. The differential pair is the physical embodiment of the excitatory/inhibitory synapse of the resistive-communication model: $G^+$ excitatory, $G^-$ inhibitory.

\subsection{Nonidealities: IR drop and the exact nodal system}

Real word/bit lines have finite resistance $r$ ($g=1/r$) per segment, degrading the delivered voltage. Let $\mathbf V^\mathrm{WL},\mathbf V^\mathrm{BL}$ be the node-voltage vectors; the exact nodal system of the passive mesh is
\begin{equation}
\mathbf Y\begin{bmatrix}\mathbf V^\mathrm{WL}\\\mathbf V^\mathrm{BL}\end{bmatrix}
=\begin{bmatrix}\mathbf i_\mathrm{src}\\\mathbf 0\end{bmatrix},
\quad
\mathbf Y=\begin{bmatrix}\mathbf Y_\mathrm{WL}+\mathbf D_r & -\mathbf G\\ -\mathbf G^{\!\top} & \mathbf Y_\mathrm{BL}+\mathbf D_c\end{bmatrix},
\label{eq:nodal}
\end{equation}
with $\mathbf D_r,\mathbf D_c$ diagonal row/column conductance sums and $\mathbf Y_\mathrm{WL},\mathbf Y_\mathrm{BL}$ tridiagonal line-conductance matrices. The ideal result \eqref{eq:vmm} is the $g\to\infty$ limit.

\begin{proposition}[IR-drop error growth]\label{prop:ir}
The relative output error of \eqref{eq:nodal} versus the ideal \eqref{eq:vmm} grows with array size as
\[
\frac{\|\mathbf I-\mathbf I_\mathrm{ideal}\|}{\|\mathbf I_\mathrm{ideal}\|}=\mathcal O\!\Bigl(\frac{N\bar G}{g}\Bigr),
\]
which bounds the practical tile size and motivates the tiling of Section~\ref{sec:system}.
\end{proposition}

\subsection{The 1T1R cell and sneak-path suppression}

In a passive ($0$T$1$R) crossbar, half-selected devices leak current along \emph{sneak paths}, corrupting read and write. Placing an access transistor in series with each memristor---the $1$T$1$R cell of Fig.~\ref{fig:synapse}---makes device $(i,j)$ conduct only when its row gate is asserted:
\begin{equation}
I_{ij}=\begin{cases}G_{ij}\,v_i,&\text{gate}_i=1,\\ \approx 0,&\text{gate}_i=0.\end{cases}
\end{equation}
The transistor also supplies the \emph{compliance current} $I_\mathrm{cc}$ during forming/set, limiting filament growth and thereby setting $G_\mathrm{max}$---the very mechanism of a precise multi-level set. The cost is area: each synapse is a transistor plus a memristor, and the differential synapse uses two.

\begin{figure}[h!]
\centering
\begin{tikzpicture}[line width=0.8pt,scale=1.0]
 \foreach \xo/\titlab/\frac in {0/{HRS (reset)}/0.25, 3.2/{forming/set}/0.7, 6.4/{LRS (set)}/1.0}{
  \begin{scope}[shift={(\xo,0)}]
    \fill[gray!55] (0,1.6) rectangle (2.0,1.9);
    \fill[gray!30] (0,-0.3) rectangle (2.0,0.0);
    \fill[blue!8] (0,0.0) rectangle (2.0,1.6);
    \draw (0,0.0) rectangle (2.0,1.9);
    \pgfmathsetmacro\h{0.05+1.5*\frac}
    \foreach \k in {0,...,7}{
      \pgfmathsetmacro\yy{0.05+0.18*\k}
      \ifdim \yy pt<\h pt
        \fill[red!75!black] (1.0,\yy) circle (0.07);
      \else
        \fill[red!75!black,opacity=0.25] (1.0,\yy) circle (0.06);
      \fi
    }
    \node[font=\scriptsize] at (1.0,2.15){\titlab};
    \node[font=\scriptsize,gray!50!black] at (1.0,1.75){top electrode};
    \node[font=\scriptsize,gray!50!black] at (1.0,-0.15){bottom electrode};
  \end{scope}
 }
 \draw[->,thick] (2.15,0.8)--(3.05,0.8) node[midway,above,font=\scriptsize]{$+V_\mathrm{set}$};
 \draw[->,thick] (5.35,0.8)--(6.25,0.8) node[midway,above,font=\scriptsize]{$I_\mathrm{cc}$};
 
 \draw[->,thick] (8.4,0.4) -- (8.6,0.4) -- (8.6,-1) 
                 -- (0.5,-1) 
                 -- (0.5,-0.4) 
                 node[pos=0,below,font=\scriptsize]{$-V_\mathrm{reset}$ (rupture)};
\end{tikzpicture}
\caption{Microscopic mechanism of bipolar resistive switching. Under $+V_\mathrm{set}$ mobile species (oxygen vacancies or metal cations, red) drift from the active electrode and nucleate a conductive filament across the oxide; the compliance current $I_\mathrm{cc}$ caps its cross-section---and hence $G_\mathrm{max}$ \eqref{eq:Gexp}. A reverse $-V_\mathrm{reset}$ ruptures the filament, returning the cell toward the HRS. Partial set/reset stops the front at intermediate heights, realizing the analog sub-levels of Section~\ref{sec:capacity}; each such event is an act of resistive communication.}
\label{fig:grid}
\end{figure}
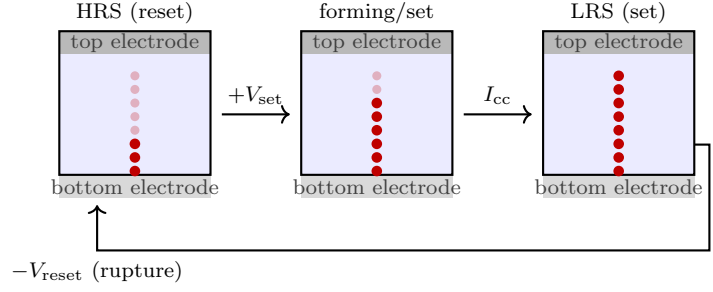

\section{On-Chip Neural-Network Architecture}\label{sec:arch}

\subsection{Layer primitive}

A single layer is one differential crossbar tile plus a row of \emph{neuron} circuits. The tile computes $\mathbf z=\mathbf W^{\!\top}\mathbf x$ as column currents (Thm.~\ref{thm:vmm}); each column current is integrated and passed through a nonlinearity $\varphi(\cdot)$:
\begin{equation}
a_j=\varphi(z_j+b_j),
\end{equation}
the bias $b_j$ injected by an extra always-on row encoding a constant. $\varphi$ may be a sigmoid (naturally produced by a subthreshold differential pair), a $\tanh$, or a rectifier (a clamped current mirror).

\begin{figure}[h!]
\centering
\begin{tikzpicture}[>=Latex,line width=0.8pt]
  \node[left,font=\footnotesize] at (-0.2,1.0){$I_\mathrm{in}$};
  \draw[->] (-0.2,1.0)--(0.5,1.0);
  \draw (0.5,1.0)--(0.5,1.2);
  \draw (0.5,1.6) pic[rotate=90]{memristor};
  \node[right,font=\footnotesize] at (0.6,1.75){$M_1$};
  \draw (0.5,2.1)--(0.5,2.5)--(1.2,2.5) node[right,font=\footnotesize]{$+V_1$};
  \draw (1.3,1.0)--(1.3,1.45); \draw (1.15,1.45)--(1.45,1.45);
  \draw (1.15,1.55)--(1.45,1.55); \draw (1.3,1.55)--(1.3,2.0);
  \node[right,font=\footnotesize] at (1.5,1.5){$C_1$};
  \draw (0.5,1.0)--(1.3,1.0);
  \draw (0.5,2.1)--(1.3,2.1) ; \draw (1.3,2.0)--(1.3,2.1);
  \draw (1.3,1.0)--(2.4,1.0);
  \node[font=\footnotesize] at (1.85,0.78){$v_1$};
  \draw (2.4,1.0)--(2.4,1.2);
  \draw (2.4,1.6) pic[rotate=90]{memristor};
  \node[right,font=\footnotesize] at (2.5,1.75){$M_2$};
  \draw (2.4,2.1)--(2.4,2.5)--(3.1,2.5) node[right,font=\footnotesize]{$-V_2$};
  \draw (3.2,1.0)--(3.2,1.45); \draw (3.05,1.45)--(3.35,1.45);
  \draw (3.05,1.55)--(3.35,1.55); \draw (3.2,1.55)--(3.2,2.0);
  \node[right,font=\footnotesize] at (3.4,1.5){$C_2$};
  \draw (2.4,1.0)--(3.2,1.0);
  \draw (2.4,2.1)--(3.2,2.1); \draw (3.2,2.0)--(3.2,2.1);
  \draw (3.2,1.0)--(4.0,1.0); \draw[->] (4.0,1.0)--(4.5,1.0);
  \node[right,font=\footnotesize] at (4.5,1.0){$v_\mathrm{out}$ (spike)};
  \draw (0.5,1.0)--(0.5,0.4); \draw (0.5,0.4)--(0.7,0.4)--(0.6,0.25)--(0.4,0.25)--(0.5,0.4);
\end{tikzpicture}
\caption{Neuristor built from two Mott memristors $M_1,M_2$, each shunted by a capacitance $C_k$ and biased with opposite polarity. Crossing the insulator--metal transition voltage triggers an all-or-nothing relaxation spike at $v_\mathrm{out}$: the electronic action potential used as the spiking neuron at each tile output.}
\label{fig:neuristor}
\end{figure}
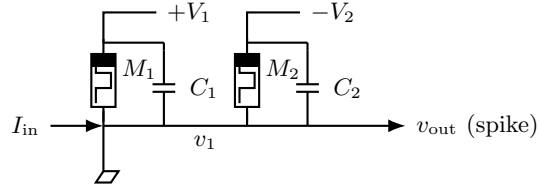

\subsection{The spiking neuristor neuron}

The threshold neuron is realized by the Mott-memristor neuristor of Fig.~\ref{fig:neuristor} \cite{pickett2013}. Each Mott channel $k\in\{1,2\}$ has a parallel capacitance $C_k$ and opposite-polarity bias; its dynamics are a FitzHugh--Nagumo-like relaxation pair,
\begin{align}
C_1\frac{dv_1}{dt} &= I_\mathrm{in} - G_1(v_1)\,v_1,\\
C_2\frac{dv_2}{dt} &= G_1(v_1)\,v_1 - G_2(v_2)\,v_2,
\end{align}
where $G_k(v)$ is the steeply nonlinear insulator--metal-transition Mott conductance. When $I_\mathrm{in}$ pushes $v_1$ past the transition voltage $V_\mathrm{th}$, the channel switches metallic, dumps charge, and relaxes---an all-or-nothing voltage spike. This is the electronic action potential of the resistive-communication model and lets the same fabric run rate-coded (continuous $\varphi$) or spike-coded \cite{indiveri2011}.

\subsection{Depth: stacking tiles}

A deep network is a stack of $K$ tiles, $\ell=1,\dots,K$,
\begin{align}
\mathbf z^{(\ell)} &= \mathbf W^{(\ell)\top}\mathbf a^{(\ell-1)} + \mathbf b^{(\ell)},\\
\mathbf a^{(\ell)} &= \varphi^{(\ell)}\!\bigl(\mathbf z^{(\ell)}\bigr),\qquad \mathbf a^{(0)}=\mathbf x.
\end{align}
The activations $\mathbf a^{(\ell)}$ are re-encoded as the row voltages of tile $\ell{+}1$. This re-encoding is exactly the \emph{resistive communication} channel: the analog state at one neuristor population's output is transmitted---by current, and ultimately by the set/reset of inter-tile coupling devices---to the next population, preserving the analog value rather than quantizing to digital. Section~\ref{sec:fabric} treats this channel quantitatively.

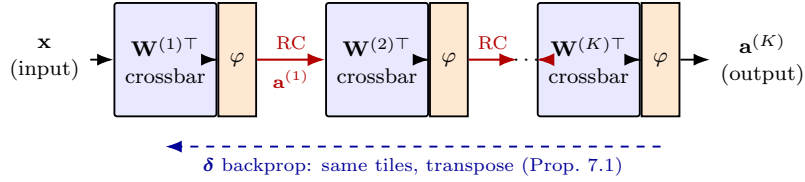
\begin{figure}[h!]
\centering
\begin{tikzpicture}[font=\scriptsize,line width=0.7pt,>=Latex,
  tile/.style={draw,rounded corners=1pt,fill=blue!9,minimum width=1.35cm,minimum height=1.5cm,align=center},
  neu/.style={draw,fill=orange!18,minimum width=0.5cm,minimum height=1.5cm,align=center},
  rc/.style={red!70!black,thick}]
  \node[align=center] (x) at (-1.4,0) {$\mathbf x$\\(input)};
  \node[tile] (w1) at (0.2,0) {$\mathbf W^{(1)\top}$\\crossbar};
  \node[neu] (n1) at (1.15,0) {$\varphi$};
  \node[tile] (w2) at (3.0,0) {$\mathbf W^{(2)\top}$\\crossbar};
  \node[neu] (n2) at (3.95,0) {$\varphi$};
  \node[tile] (wk) at (5.8,0) {$\mathbf W^{(K)\top}$\\crossbar};
  \node[neu] (nk) at (6.75,0) {$\varphi$};
  \node[align=center] (y) at (8.1,0) {$\mathbf a^{(K)}$\\(output)};
  \draw[->] (x)--(w1);
  \draw[->] (w1)--(n1);
  \draw[rc,->] (n1) -- node[above]{\tiny RC} node[below]{\tiny $\mathbf a^{(1)}$} (w2);
  \draw[->] (w2)--(n2);
  \draw[rc,->] (n2) -- node[above]{\tiny RC} (4.85,0);
  \node at (5.0,0) {$\cdots$};
  \draw[rc,->] (5.15,0) -- (wk);
  \draw[->] (wk)--(nk);
  \draw[->] (nk)--(y);
  \draw[blue!60!black,dashed,->] (6.75,-1.15) -- node[below]{\tiny $\bm\delta$ backprop: same tiles, transpose (Prop.~\ref{prop:transpose})} (0.2,-1.15);
\end{tikzpicture}
\caption{Multilayer pipeline. Each layer is a crossbar tile ($\mathbf W^{(\ell)\top}$) followed by a neuron row ($\varphi$); activations propagate left-to-right over the active resistive-communication (RC) channels of Section~\ref{sec:fabric}, staying analog between tiles. The dashed return path is the backward pass, executed on the \emph{same} tiles driven in transpose (Proposition~\ref{prop:transpose}).}
\label{fig:pipeline}
\end{figure}

\subsection{Why multi-level matters here}

If each device held a single bit, representing a weight to $b$ bits would need $b$ devices and a digital adder---reintroducing the von Neumann cost we set out to remove. With $L\!\approx\!2^b$ analog sub-levels per device, one differential pair holds the entire weight and the multiply is a single Ohm's-law event. Density and energy efficiency scale directly with $\log_2 L$; the entire architecture rests on the many-sub-level capability of Section~\ref{sec:capacity}.

\section{In-Situ Learning on the Conductance Lattice}\label{sec:learning}

\subsection{Backpropagation, physically}

Training minimizes a loss $\mathcal L$. The gradient at layer $\ell$ is the outer product of the layer input with the back-propagated error,
\begin{equation}
\frac{\partial \mathcal L}{\partial \mathbf W^{(\ell)}}=\mathbf a^{(\ell-1)}\bm\delta^{(\ell)\top},
\quad \bm\delta^{(\ell)}=\varphi'\!\bigl(\mathbf z^{(\ell)}\bigr)\odot\bigl(\mathbf W^{(\ell+1)}\bm\delta^{(\ell+1)}\bigr).
\label{eq:bp}
\end{equation}

\begin{proposition}[Free transpose]\label{prop:transpose}
The crossbar that computes the forward product $\mathbf W^{\!\top}\mathbf a$ also computes the backward product $\mathbf W\bm\delta$ when driven from its columns and read from its rows. Both passes of backpropagation are therefore analog VMMs on identical hardware.
\end{proposition}

\subsection{The outer-product update}

The update $\Delta W_{ij}=-\eta\,a_i\,\delta_j$ is an outer product, realized by a pulse on row $i$ of amplitude/duration $\propto a_i$ \emph{coincident} with a pulse on column $j$ $\propto\delta_j$; only the intersection device sees the full overdrive, and by the steep $\sinh$ law only it changes state:
\begin{equation}
\Delta x_{ij}\propto\sinh\!\Bigl(\tfrac{v_i^\mathrm{row}+v_j^\mathrm{col}}{V_0}\Bigr)-\sinh\!\Bigl(\tfrac{v_i^\mathrm{row}}{V_0}\Bigr)-\sinh\!\Bigl(\tfrac{v_j^\mathrm{col}}{V_0}\Bigr).
\end{equation}
Coincident pulses give a multiplicative (Hebbian) update while half-selected devices are nearly untouched---exactly spike-timing-dependent plasticity realized by overlapping pre/post pulses \cite{prezioso2015}. The resistive-communication ``set/reset that remembers past memristance levels'' is the physical substrate of learning.

\subsection{Discretization onto a finite lattice}

The achievable conductances form a fine but finite lattice $\{G_1,\dots,G_L\}$. Writing the realized weight as the nearest lattice point $\hat W=Q(W)$, the update becomes a stochastic-rounding step
\begin{equation}
W\leftarrow Q\bigl(W-\eta\,a\,\delta+\xi\bigr),
\end{equation}
$\xi$ the program-and-verify residual. For large $L$ the quantization acts as additive noise of variance $\sigma_Q^2=(\Delta G/\beta)^2/12$; stochastic gradient descent is robust to it provided
\begin{equation}
\eta\,\|a\,\delta\|\gtrsim\Delta W=\beta^{-1}\Delta G,
\end{equation}
i.e.\ the learning step is at least one lattice spacing. Many sub-levels (small $\Delta G$) lower the smallest trainable update and hence the accuracy floor---another reason density matters.

\subsection{Asymmetric nonlinearity and its correction}

Set (potentiation) and reset (depression) are asymmetric and nonlinear in pulse number $n$:
\begin{equation}
G_\uparrow(n)=G_\mathrm{min}+(G_\mathrm{max}-G_\mathrm{min})\frac{1-e^{-n/\tau_p}}{1-e^{-n_\mathrm{max}/\tau_p}},
\end{equation}
with a different time constant for $G_\downarrow$. Uncorrected, this biases learning. We linearize by inverting the curve: to move from $G_a$ to $G_b$,
\begin{equation}
\Delta n=\tau_p\ln\frac{G_\mathrm{max}-G_a}{G_\mathrm{max}-G_b},
\end{equation}
so equal weight steps map to equal conductance steps. With program-and-verify this restores a near-ideal, symmetric update suitable for gradient descent \cite{ambrogio2018}.

\section{Mixed-Signal Peripherals}\label{sec:periph}

The analog core is bounded by data converters whose precision must match the cell. Inputs are applied by a DAC, preferably as \emph{pulse-width} encoding, which keeps every device at a single read voltage $V_r$ (avoiding the nonlinearity of \eqref{eq:Gexp}) and encodes operand $x_i$ in the time $\tau_i=x_i\tau_\mathrm{max}$:
\begin{equation}
q_j=\int_0^{\tau_\mathrm{max}}\!\! I_j(t)\,dt=V_r\sum_i G_{ij}\tau_i=V_r\tau_\mathrm{max}\sum_i G_{ij}x_i.
\end{equation}
The accumulated charge $q_j$ is then digitized. To not waste the cell's $\log_2 L$ bits, the column ADC must resolve at least
\begin{equation}
b_\mathrm{ADC}\ge\log_2\!\Bigl(N\tfrac{G_\mathrm{max}}{\sigma_G}\Bigr)\approx\log_2 N+\log_2 L,
\end{equation}
since up to $N$ devices sum onto a column. The $\log_2 N$ fan-in term is the main reason large layers are partitioned: smaller $N$ relaxes the ADC and keeps SAR-converter energy ($\sim\!2^{b_\mathrm{ADC}}$) under control. Converter energy, not device energy, dominates the system budget, so peripheral precision is co-designed with tile size and cell level count.

\subsection{Convolutional mapping}

A convolution with kernel $\mathbf K\in\mathbb R^{C_\mathrm{in}\times k\times k}$ producing $C_\mathrm{out}$ channels is lowered by the im2col/Toeplitz transform: the kernel becomes a weight matrix $\mathbf W\in\mathbb R^{(C_\mathrm{in}k^2)\times C_\mathrm{out}}$ programmed once, while image patches $\mathbf p_t$ stream as row inputs, $\mathbf y_t=\mathbf W^{\!\top}\mathbf p_t$. Weight reuse across all $H'W'$ spatial positions is free---the kernel is written once and amortized over every patch, precisely where in-memory computing beats a digital accelerator that re-fetches weights \cite{yao2020}.

\section{Energy and Area Model}\label{sec:energy}

\subsection{Energy per MAC}

A multiply at junction $(i,j)$ over a read pulse $\tau_r$ at bias $V_r$ dissipates $E_\mathrm{MAC}=G_{ij}V_r^2\tau_r$. For an $N\times M$ tile doing $NM$ MACs per read, the average conductance $\bar G$ sets $E_\mathrm{tile}/NM=\bar G V_r^2\tau_r$. With $\bar G\!\sim\!50~\mu$S, $V_r=0.2$~V, $\tau_r=10$~ns this is $E_\mathrm{MAC}\approx2\times10^{-14}$~J, i.e.\ a few tens of femtojoules per MAC---orders of magnitude below digital MAC energy, because no data is moved.

\begin{proposition}[Energy--resolution trade]\label{prop:energy}
To resolve $L$ levels at margin $m$, \eqref{eq:Lth} forces a minimum read budget $V_r^2\tau_r$, giving
\[
E_\mathrm{MAC}\gtrsim m^2 k_B T\,\frac{(L-1)^2}{\mathrm{SNR\ budget}},
\]
the analog analogue of a Landauer-style bound: more sub-levels cost more read energy, quadratically.
\end{proposition}

\subsection{Area and density}

The $1$T$1$R differential synapse occupies roughly $2(A_T+A_M)$, dominated by the access transistors; the memristor sits in the back-end-of-line above the transistor at little extra footprint. Storing $b=\log_2 L$ bits per device gives bit density
\begin{equation}
\rho_\mathrm{bit}=\frac{\log_2 L}{2(A_T+A_M)},
\end{equation}
which for $L$ in the hundreds beats SRAM weight storage by one to two orders of magnitude---the quantitative payoff of the multi-level cell.

\section{Resistive Communication Between Tiles}\label{sec:fabric}

A network larger than one tile must move analog activations between tiles without collapsing them to digital. We model the inter-tile channel as a chain of neuristors whose set/reset events carry the activation forward---the resistive communication of \cite{trejo2017}, now quantitative.

\subsection{The channel}

Treat the link as a memristive transmission line: a ladder of series memristances $M_k$ and shunt capacitances $C_k$. A small-signal activation $a(t)$ launched at one end propagates as a damped wave,
\begin{equation}
\frac{\partial^2 v}{\partial t^2}=\frac{1}{MC}\frac{\partial^2 v}{\partial \xi^2}-\frac{1}{\tau_M}\frac{\partial v}{\partial t},
\end{equation}
whose group velocity sets inter-tile latency and damping $1/\tau_M$ sets attenuation. Biasing the neuristors just below their firing threshold makes the line \emph{active}: each stage regenerates the pulse (the Mott relaxation oscillator supplies gain), giving soliton-like, attenuation-free propagation---the electronic analogue of saltatory conduction along an axon.

\subsection{A nanoscale protocol}

To interoperate with other nanodevices the channel follows the IEEE~1906.1 framework \cite{ieee1906} for nanoscale communication: the activation is the \emph{message}, the neuristor chain the \emph{microscopic channel}, and the column-current readout the \emph{specificity} component. Where a molecular-communication channel obeys Fick's diffusion law,
\begin{equation}
c(r,t)=\frac{Q}{(4\pi D t)^{3/2}}\,e^{-r^2/4Dt},
\end{equation}
the resistive channel replaces diffusing molecules with drifting ions and propagating filament fronts, trading the slow $\sqrt{t}$ diffusion spread for the fast, regenerative wave above---orders of magnitude faster while remaining a genuine nanoscale communication primitive.

\section{System-Level Considerations}\label{sec:system}

\subsection{Mapping a network onto tiles}

A layer with $N$ inputs and $M$ outputs needs an $N\times 2M$ differential tile. Layers larger than the maximum reliable tile (bounded by Prop.~\ref{prop:ir}) are partitioned across tiles, their partial column currents summed in the analog domain before the neuron. Convolutional layers map by the Toeplitz unrolling of Section~\ref{sec:periph}.

\subsection{Calibration and chip-in-the-loop training}

Device-to-device variability is handled by \emph{chip-in-the-loop} training: the forward pass runs on the physical array, the loss is evaluated, and the program-and-verify controller applies the analog outer-product update of Section~\ref{sec:learning}. Because errors are measured on real hardware, the learned weights absorb fixed nonidealities (offset, gain, fixed sneak paths), a key advantage of in-situ analog learning over train-then-transfer \cite{alibart2013,burr2015}.

\subsection{Reliability}

Endurance ($\sim\!10^{6}$--$10^{9}$ cycles) limits how often weights may be updated; inference, being read-only, is effectively unlimited. The differential pair, periodic refresh, and drift-aware level spacing of Section~\ref{sec:capacity} keep the network within its accuracy budget over its operating life.

\section{A Worked Example: A Three-Layer Classifier}\label{sec:example}

To make the preceding analysis concrete we size a small but complete classifier, a $784\!-\!256\!-\!128\!-\!10$ fully-connected network (an MNIST-scale topology), entirely on the proposed fabric.

\subsection{Tiling}

With a maximum reliable tile of $256\times256$ (set by Prop.~\ref{prop:ir} with $\bar G=50~\mu$S, $g=1$~mS per segment, target error $<1\%$), the layers map as in Table~\ref{tab:map}. Each weight uses a differential pair, so a layer with $M$ outputs needs $2M$ columns.

\begin{table}[h!]
\centering
\caption{Mapping the $784\!-\!256\!-\!128\!-\!10$ network onto $256\times256$ tiles.}\label{tab:map}
\vspace{0.15cm}
\begin{tabular}{@{}lcccc@{}}
\toprule
\textbf{Layer} & \textbf{In$\times$Out} & \textbf{Diff. cols} & \textbf{Row tiles} & \textbf{Col tiles} \\
\midrule
FC1 & $784\times256$ & $512$ & $\lceil784/256\rceil=4$ & $\lceil512/256\rceil=2$ \\
FC2 & $256\times128$ & $256$ & $1$ & $1$ \\
FC3 & $128\times10$  & $20$  & $1$ & $1$ \\
\bottomrule
\end{tabular}
\end{table}

FC1 needs $4\times2=8$ tiles whose partial column currents are summed in the analog domain (Section~\ref{sec:system}); FC2 and FC3 fit in one tile each. The total is $10$ tiles, $\approx\!2.7\times10^{5}$ differential synapses, i.e.\ $\approx\!5.4\times10^{5}$ $1$T$1$R cells.

\subsection{Precision budget}

Each cell stores $\log_2 L\approx 8$ bits (Table~\ref{tab:budget}). The fan-in of FC1 is $N=784$, partitioned into four row-tiles of $\le256$, so each column ADC resolves $b_\mathrm{ADC}\ge\log_2 256+\log_2 L\approx 8+8=16$ bits before partial-sum accumulation; after the four partial sums combine digitally the effective accumulation precision is maintained. FC2 ($N=256$) and FC3 ($N=128$) require $16$ and $15$ bits respectively. These are demanding but within reach of pipelined SAR/$\Sigma\Delta$ converters time-shared across columns.

\subsection{Throughput and energy}

One inference is three analog VMMs (Thm.~\ref{thm:vmm}), each completing in one read window $\tau_r=10$~ns plus conversion. Pipelined across the three layers, the steady-state throughput is one classification per $\sim\!\tau_\mathrm{ADC}$. The dynamic energy is dominated by reads and conversions:
\begin{equation}
E_\mathrm{inf}\approx \underbrace{N_\mathrm{cell}\,\bar G V_r^2\tau_r}_{\text{device}}+\underbrace{N_\mathrm{ADC}\,E_\mathrm{ADC}}_{\text{conversion}}.
\end{equation}
With $N_\mathrm{cell}\approx5.4\times10^{5}$, the device term is $5.4\times10^{5}\cdot 50~\mu\mathrm{S}\cdot(0.2\,\mathrm V)^2\cdot10\,\mathrm{ns}\approx 11$~nJ. Per Prop.~\ref{prop:energy} the converters dominate; even so the total stays in the tens-of-nJ-per-inference range, two to three orders of magnitude below a digital baseline that must fetch $\sim\!2.7\times10^{5}$ weights from SRAM per inference.

\section{Comparison with Digital Accelerators}\label{sec:comparison}

Table~\ref{tab:compare} summarizes the structural differences. The decisive factors are that (i) weights never move---they are the computation---eliminating the dominant data-movement energy, and (ii) the multi-level cell collapses $b$ digital storage cells plus an adder tree into a single Ohm's-law event, so density and energy improve by the factor $\log_2 L$ relative to a binary memristive design.

\begin{table}[h!]
\centering
\caption{Analog multi-level in-memory fabric vs.\ a digital MAC array.}\label{tab:compare}
\vspace{0.15cm}
\begin{tabular}{@{}p{3.4cm}p{4.4cm}p{4.4cm}@{}}
\toprule
\textbf{Aspect} & \textbf{Digital MAC array} & \textbf{This work (analog IMC)} \\
\midrule
Weight storage & SRAM/DRAM, off the datapath & In the synapse itself ($\log_2 L$ bits/cell) \\
MAC mechanism & Clocked arithmetic units & Ohm + Kirchhoff, \eqref{eq:vmm} \\
VMM time & $O(NM/P)$, $P$ lanes & $O(1)$ per tile (Thm.~\ref{thm:vmm}) \\
Dominant energy & Data movement & ADC/DAC conversion (Prop.~\ref{prop:energy}) \\
Precision & Exact (digital) & Analog, noise-bounded \eqref{eq:Lth} \\
Training & Off-chip & In-situ, free transpose (Prop.~\ref{prop:transpose}) \\
Main limiter & Memory bandwidth & Device variability, IR drop (Prop.~\ref{prop:ir}) \\
\bottomrule
\end{tabular}
\end{table}

The analog approach is not universally superior: it trades exactness for efficiency and is best suited to inference-dominated, error-tolerant workloads---exactly the regime of neural-network deployment. The multi-level memristor is what makes the trade favorable, because without many sub-levels the density advantage collapses and the architecture reduces to an ordinary, area-hungry binary crossbar.

\section{Computing Large Language Models on the Multi-Level Fabric}\label{sec:llm}

Large language models (LLMs) are the most demanding deployment target of contemporary computing, and they are also the workload for which an analog multi-level in-memory fabric is most advantageous, because their cost is overwhelmingly dominated by dense matrix multiplications against \emph{static} weights---precisely the operation the crossbar performs for free.

\subsection{Decomposition of the transformer workload}

A decoder-only transformer of $n_\mathrm{layer}$ blocks, model width $d$, feed-forward width $d_\mathrm{ff}=4d$, and vocabulary $V$ maps onto the fabric as a set of weight matrices, each programmed once into crossbar tiles and reused for every token. Per layer the weight-bearing operations are
\begin{align}
\text{QKV proj.:}\;&\ \mathbf W_{Q},\mathbf W_{K},\mathbf W_{V}\in\mathbb R^{d\times d},\\
\text{output proj.:}\;&\ \mathbf W_{O}\in\mathbb R^{d\times d},\\
\text{FFN:}\;&\ \mathbf W_{1}\in\mathbb R^{d\times d_\mathrm{ff}},\ \mathbf W_{2}\in\mathbb R^{d_\mathrm{ff}\times d}.
\end{align}
The total static parameter count is
\begin{equation}
P \approx n_\mathrm{layer}\,(4d^2 + 2\,d\,d_\mathrm{ff}) = 12\,n_\mathrm{layer}\,d^2,
\end{equation}
and the per-token inference cost (MACs) of the linear maps is $2P$. For a $7$-billion-parameter model this is $\sim\!1.4\times10^{10}$ MACs per generated token---every one of which, on the proposed fabric, is a single Ohm's-law event drawn from a weight already resident in a memristor.

\subsection{Mapping weights to multi-level tiles}

Each weight element occupies one differential pair storing $b=\log_2 L$ bits (Section~\ref{sec:capacity}). The number of $1$T$1$R cells required to hold the entire model is
\begin{equation}
N_\mathrm{cell} = 2P,
\end{equation}
independent of the arithmetic precision of a digital datapath, because the precision lives \emph{in the analog level count}, not in extra storage. The number of $256\times256$ tiles is $N_\mathrm{tile}=\lceil P/256^2\rceil$ (differential columns folded in), and the QKV/output/FFN matrices stream their token activations as row voltages exactly as in Theorem~\ref{thm:vmm}. The attention score $\mathbf S=\mathbf Q\mathbf K^{\!\top}/\sqrt{d_k}$ and the context $\mathbf S\mathbf V$ are \emph{dynamic} (data$\times$data) products; these are handled either by a small auxiliary digital unit or by transiently writing $\mathbf K,\mathbf V$ into a scratch crossbar---the KV cache becomes a literal resistive memory, with each cached key/value an analog conductance.

\subsection{Energy per token: an analog roofline}

The dynamic energy to generate one token is the read energy of every weight plus the conversion overhead:
\begin{equation}
E_\mathrm{tok} \approx \underbrace{2P\,\bar G V_r^2 \tau_r}_{\text{analog MAC}} + \underbrace{N_\mathrm{ADC}^\mathrm{(tok)}\,E_\mathrm{ADC}}_{\text{conversion}}.
\end{equation}
With $\bar G=50~\mu$S, $V_r=0.2$~V, $\tau_r=10$~ns the analog term is $\sim\!2\times10^{-14}$~J per MAC, i.e.\ $2P\,E_\mathrm{MAC}\approx0.3$~mJ per token for a $7$B model---before conversion. Crucially the weights are \emph{never fetched}: in a digital accelerator the same token requires reading $P$ weights ($\sim\!14$~GB at INT16) from DRAM, whose energy at $\sim\!10$~pJ/byte dwarfs the arithmetic by two to three orders of magnitude. The transformer is memory-bandwidth bound on every platform that stores weights off the datapath; the multi-level fabric removes that bound by construction.

\begin{principle}[Why LLMs in particular]\label{prin:llm}
Autoregressive generation reuses the entire weight set once per token with batch size one, so arithmetic intensity is low and weight \emph{movement}, not weight \emph{multiplication}, dominates. An in-memory fabric makes movement zero; a multi-level cell makes the resident footprint minimal. The two effects compound precisely in the regime where conventional hardware is worst.
\end{principle}

\subsection{Advantage over binary/few-level ReRAM}

A ReRAM accelerator that stores only $1$--$2$ bits per cell must represent a $b$-bit weight across $\lceil b/b_\mathrm{cell}\rceil$ devices and recombine them with shift-and-add peripherals. Relative to such a design the multi-level fabric improves the three figures of merit by the level-count factor:
\begin{align}
\frac{\text{area}_\mathrm{binary}}{\text{area}_\mathrm{ML}} &\approx \frac{b}{b_\mathrm{cell}} = \log_2 L \ \ (\text{for }b_\mathrm{cell}=1),\\
\frac{E_\mathrm{binary}}{E_\mathrm{ML}} &\approx \log_2 L \times \frac{1}{1-\eta_\mathrm{SA}},
\end{align}
where $\eta_\mathrm{SA}$ is the shift-and-add overhead avoided when a single analog read already yields the full-precision product. For $L\!\sim\!256$ this is an $8\times$ density and energy advantage from the cell alone, on top of the shared in-memory benefit. The penalty---analog noise---is bounded by Theorem~\ref{thm:levels} and absorbed by chip-in-the-loop training (Section~\ref{sec:system}); LLM inference is known to tolerate aggressive weight quantization, so the $\sim\!8$ effective bits of a multi-level cell are comfortably sufficient.

\subsection{Advantage over traditional CMOS}

Against a digital CMOS accelerator the advantage is structural rather than incremental, and is summarized in Table~\ref{tab:llm}. The CMOS datapath pays (i) DRAM/SRAM weight-fetch energy on every token, (ii) clocked MAC energy, and (iii) the area of a deep SRAM hierarchy; the analog fabric pays none of these, trading them for converter energy and bounded analog noise. The decisive scaling is in the weight-movement term, which vanishes.

\begin{table}[h!]
\centering
\caption{Per-token cost of an $L$-layer LLM on three substrates (scaling, weight-related terms).}\label{tab:llm}
\vspace{0.15cm}
\begin{tabular}{@{}p{3.0cm}p{3.4cm}p{3.4cm}p{2.4cm}@{}}
\toprule
\textbf{Term} & \textbf{CMOS digital} & \textbf{Binary ReRAM} & \textbf{This work} \\
\midrule
Weight movement & $O(P)$ DRAM reads & none (in-memory) & none \\
MAC energy & clocked, $O(P)$ & analog $\times\log_2 L$ cells & analog, $O(P)$ \\
Storage area & SRAM, $O(Pb)$ & $O(P\log_2 L)$ cells & $O(P)$ cells \\
Bits per cell & --- & $1$--$2$ & $\log_2 L\ (\sim\!8)$ \\
Dominant cost & memory bandwidth & shift-and-add + ADC & ADC \\
\bottomrule
\end{tabular}
\end{table}

\subsection{System organization for a full model}

A practical deployment pipelines the $n_\mathrm{layer}$ transformer blocks across physical tile clusters, with the resistive-communication channel of Section~\ref{sec:fabric} carrying each block's activation vector to the next as a regenerated analog pulse train rather than a digitized bus transaction. Weights are written once at load time (amortizing the endurance cost of Section~\ref{sec:system}), residency is permanent and non-volatile, and the chip draws static power only through leakage---so an idle model consumes essentially nothing, unlike a DRAM-resident model that must be continuously refreshed. The combination of permanent multi-level weight residency, zero weight movement, and analog VMM is what makes a single-chip, low-power LLM plausible on this fabric.

\section{In-Memory Self-Attention}\label{sec:attention}

The one part of a transformer that is \emph{not} a static-weight product is self-attention,
\newline
$\mathbf A=\mathrm{softmax}(\mathbf Q\mathbf K^{\!\top}/\sqrt{d_k})\,\mathbf V$, whose two matrix products are data$\times$data. Section~\ref{sec:llm} deferred these to a digital unit; here we show they too can be computed in the analog domain, which is what makes the fabric a \emph{complete} transformer engine rather than a weight-product accelerator.

\subsection{Dynamic crossbars and volatile scratch}

The projections $\mathbf Q,\mathbf K,\mathbf V$ are produced by static-weight VMMs
(Section~\ref{sec:llm}). To form the score $\mathbf S=\mathbf Q\mathbf K^{\!\top}$
we \emph{transiently} program the $d_k$ key vectors as the columns of a scratch
crossbar, $G^{K}_{ij}\propto K_{ji}$, and drive the query as row voltages; the
column currents are
\begin{equation}
S_{tj}=\sum_{i} G^{K}_{ij}\,Q_{ti} \;\propto\; (\mathbf Q\mathbf K^{\!\top})_{tj},
\end{equation}
one analog VMM. The context $\mathbf A\mathbf V$ is a second VMM with $\mathbf V$
written as conductances. The defining requirement---fast write, short
retention---is met not by the non-volatile synapse but by a \emph{volatile},
diffusive memristor whose state relaxes on the millisecond scale of the KV-cache
lifetime. The same materials family supplies both: an engineered low-barrier
$U_a$ (Section~\ref{sec:materials}) gives a self-erasing scratch cell, so the KV
cache is literally a decaying conductance pattern. This sidesteps the endurance
cost of rewriting non-volatile cells every token (Section~\ref{sec:system}).

\subsection{Analog softmax by the translinear principle}

The normalization is performed in the current domain without leaving the analog
core. Exponentiation is the native law of a subthreshold MOS (or bipolar) device,
$I=I_0 e^{\kappa V/V_T}$, so applying each score $S_{tj}$ as a gate voltage yields
currents $I_j\propto e^{S_{tj}}$. Kirchhoff's law sums them on a shared node,
$\sum_j I_j$, and a translinear normalizer divides, giving exactly
\begin{equation}
A_{tj}=\frac{e^{S_{tj}}}{\sum_{k} e^{S_{tk}}},
\end{equation}
the softmax, as a ratio of currents. The temperature $1/\sqrt{d_k}$ is set by a
single bias. Thus the entire attention operator---two matrix products and a
softmax---executes as analog VMM, current-mode exponentiation, and KCL
normalization, with no digital matrix arithmetic on the critical path.

\begin{theorem}[Attention in constant tile-passes]\label{thm:attn}
For a sequence of length $n_s$ and head dimension $d_k$, in-memory self-attention
requires $O(1)$ analog tile-passes per query (independent of $n_s$ within a tile of
$n_s$ columns), $O(n_s d_k)$ scratch devices, and energy
$E_\mathrm{attn}=O(n_s d_k)\,E_\mathrm{MAC}$, versus the $O(n_s d_k)$ clocked MACs
and the $O(n_s d_k)$ memory traffic of a digital implementation.
\end{theorem}

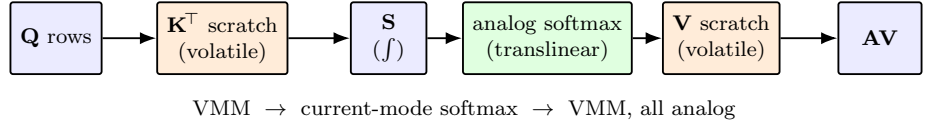
\begin{figure}[h!]
\centering
\begin{tikzpicture}[font=\scriptsize,line width=0.7pt,>=Latex,
  blk/.style={draw,rounded corners=1pt,minimum height=0.95cm,align=center,fill=blue!8}]
  \node[blk,minimum width=1.2cm] (q) at (0,0){$\mathbf Q$ rows};
  \node[blk,fill=orange!15,minimum width=1.7cm] (kk) at (2.2,0){$\mathbf K^{\!\top}$ scratch\\(volatile)};
  \node[blk,minimum width=1.0cm] (s) at (4.4,0){$\mathbf S$\\($\int$)};
  \node[blk,fill=green!12,minimum width=1.7cm] (sm) at (6.5,0){analog softmax\\(translinear)};
  \node[blk,fill=orange!15,minimum width=1.4cm] (vv) at (8.8,0){$\mathbf V$ scratch\\(volatile)};
  \node[blk,minimum width=1.1cm] (o) at (10.9,0){$\mathbf A\mathbf V$};
  \draw[->] (q)--(kk); \draw[->] (kk)--(s); \draw[->] (s)--(sm);
  \draw[->] (sm)--(vv); \draw[->] (vv)--(o);
  \node[align=center] at (5.4,-0.95){\scriptsize VMM \;$\to$\; current-mode softmax \;$\to$\; VMM, all analog};
\end{tikzpicture}
\caption{In-memory self-attention. Queries drive a volatile scratch crossbar
holding $\mathbf K^{\!\top}$ to produce scores $\mathbf S$ (one VMM); a current-mode
translinear stage computes the softmax; a second VMM against a volatile $\mathbf V$
crossbar yields the context $\mathbf A\mathbf V$. Diffusive (self-erasing)
memristors serve as the KV scratch, avoiding non-volatile rewrite wear.}
\label{fig:attention}
\end{figure}

\section{CMOS Integration: FEOL, BEOL, and Foundry Compatibility}\label{sec:cmos}

The architecture is only useful if it can be built in the standard CMOS flow that
every foundry already runs. It can: the memristor of Principle~\ref{prin:material}
is a back-end-of-line (BEOL) element built from materials (HfO$_2$, TiN, Ti) that
are already qualified, while the access transistors, neurons, converters, and
controller are ordinary front-end-of-line (FEOL) CMOS. This is the key practical
claim of the paper, and it is what makes a \emph{memristor-based FPGA} a near-term
device rather than a new fabrication science.

\subsection{FEOL: the transistors}

The front end is unmodified. The $1$T$1$R access transistors
(Fig.~\ref{fig:synapse}a), the transimpedance/integrator neurons, the
pulse-width DACs, the SAR/$\Sigma\Delta$ column ADCs, and the digital
configuration logic are all standard logic and analog cells in a baseline node
(e.g.\ $28$ or $22$~nm). The access transistor is sized to pass the set
compliance current $I_\mathrm{cc}$ (typically $10$--$100~\mu$A), which is modest;
its $W/L$ is the only synapse-specific FEOL constraint. No exotic implants, no
new well structures, no changes to the transistor module are required.

\subsection{BEOL: inserting the memristor}

The memristor is fabricated \emph{between two existing metal layers} during the
back end, after the transistors and the lower interconnect are complete. A
typical insertion sits between metal-$M_4$ and $M_5$:
\begin{enumerate}[label=(\roman*)]
\item pattern and fill the $M_4$ via; deposit and CMP the TiN bottom electrode;
\item ALD the HfO$_{2-x}$ switching layer ($3$--$10$~nm) at $\le 400^{\circ}$C---within the BEOL thermal budget;
\item sputter the Ti oxygen-scavenging layer and the TiN top electrode;
\item etch the MIM pillar, encapsulate, and continue with the $M_5$ via and the remaining interconnect/passivation.
\end{enumerate}
Every step uses tools already in the line (ALD, PVD, CMP, reactive-ion etch); the
only additions are two masks and the HfO$_2$/Ti/TiN depositions, all under the
$\sim\!400^{\circ}$C ceiling that protects the FEOL silicide and the lower copper.

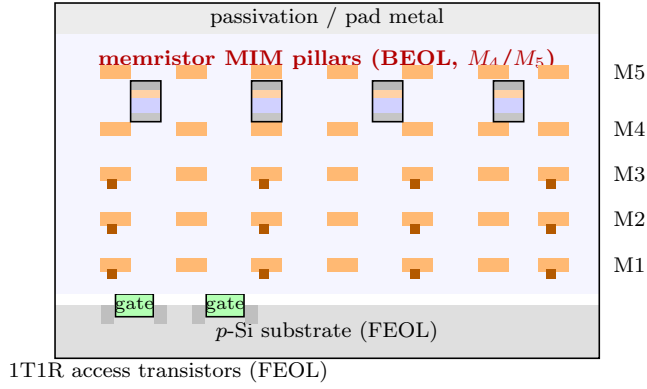
\begin{figure}[h!]
\centering
\begin{tikzpicture}[font=\scriptsize,line width=0.6pt]
  \def\w{7.2}
  \fill[gray!25] (0,0) rectangle (\w,0.7); \node at (\w/2,0.35){$p$-Si substrate (FEOL)};
  \foreach \x in {0.8,2.0}{
    \fill[green!30] (\x,0.55) rectangle (\x+0.5,0.85);
    \draw (\x,0.55) rectangle (\x+0.5,0.85);
    \node at (\x+0.25,0.7){\tiny gate};
    \fill[gray!50] (\x-0.18,0.45) rectangle (\x-0.02,0.7); 
    \fill[gray!50] (\x+0.52,0.45) rectangle (\x+0.68,0.7);
  }
  \node[align=center] at (1.5,-0.18){\scriptsize 1T1R access transistors (FEOL)};
  \fill[blue!4] (0,0.85) rectangle (\w,4.3);
  \foreach \y/\lab in {1.15/M1,1.75/M2,2.35/M3}{
    \foreach \x in {0.6,1.6,2.6,3.6,4.6,5.6,6.4}{ \fill[orange!55] (\x,\y) rectangle (\x+0.4,\y+0.18);}
    \node[right] at (\w+0.05,\y+0.09){\lab};
  }
  \foreach \y in {1.05,1.65,2.25}{ \foreach \x in {0.7,2.7,4.7,6.5}{ \fill[orange!70!black] (\x,\y) rectangle (\x+0.12,\y+0.12);}}
  \foreach \x in {0.6,1.6,2.6,3.6,4.6,5.6,6.4}{ \fill[orange!55] (\x,2.95) rectangle (\x+0.4,3.13);}
  \node[right] at (\w+0.05,3.04){M4};
  \foreach \x in {1.0,2.6,4.2,5.8}{
    \fill[gray!45] (\x,3.13) rectangle (\x+0.4,3.25);     
    \fill[blue!18] (\x,3.25) rectangle (\x+0.4,3.45);     
    \fill[orange!35](\x,3.45) rectangle (\x+0.4,3.55);    
    \fill[gray!55] (\x,3.55) rectangle (\x+0.4,3.67);     
    \draw (\x,3.13) rectangle (\x+0.4,3.67);
  }
  \node[red!70!black,align=center] at (\w/2,3.95){\scriptsize \textbf{memristor MIM pillars (BEOL, $M_4$/$M_5$)}};
  \foreach \x in {0.6,1.6,2.6,3.6,4.6,5.6,6.4}{ \fill[orange!55] (\x,3.7) rectangle (\x+0.4,3.88);}
  \node[right] at (\w+0.05,3.79){M5};
  \fill[gray!15] (0,4.3) rectangle (\w,4.7); \node at (\w/2,4.5){passivation / pad metal};
  \draw (0,0) rectangle (\w,4.7);
\end{tikzpicture}
\caption{Schematic cross-section of the integrated chip. FEOL transistors and the
lower copper interconnect ($M_1$--$M_4$) are an unmodified logic process; the
memristive synapses are inserted as MIM pillars in the BEOL between $M_4$ and
$M_5$, then buried under the upper metal and passivation. The crossbar word/bit
lines are realized in $M_4$/$M_5$ directly above the access transistors,
giving a compact $1$T$1$R footprint.}
\label{fig:beol}
\end{figure}

\begin{principle}[Foundry readiness]\label{prin:foundry}
Because the memristor is a BEOL module of pre-qualified materials inserted between
existing metals, the design is a \emph{more-than-Moore} add-on to a baseline node,
not a new technology. The same reasoning underlies commercial embedded-ReRAM
offerings already in production at major foundries; the contribution here is the
\emph{analog multi-level} operating point and the FPGA-like reconfigurable fabric
built on top of it.
\end{principle}

\section{SPICE Simulation}\label{sec:spice}

We validate the device and the VMM primitive with a SPICE model implementing the
compact equations of Section~\ref{sec:compact}. Listing~\ref{lst:spice} gives a
self-contained \texttt{ngspice} subcircuit: a behavioural state variable $x$ is
held on a unit capacitor at node \texttt{x}, integrated from the thresholded drift
current of \eqref{eq:compactstate} with a Biolek-type window, and the terminal
current is the HRS/LRS interpolation. A second deck wires four such devices into a
$2\times2$ crossbar and reads the column currents to confirm
$\mathbf I=\mathbf G^{\!\top}\mathbf v$.

\begin{lstlisting}[caption={ngspice memristor subcircuit and crossbar test.},label={lst:spice}]
* ---- multilevel memristor (compact model, Sec. VI) ----
.subckt memr p n PARAMS: x0=0.1
* state variable on node x (1F integrator), bounded [0,1]
Cx x 0 1
.func win(xx,vv) {1 - pow((2*xx-1),2*2)} ; Biolek window (p=2)
.func drift(vv) {sgn(vv)*pow(max(abs(vv)-0.6,0),2)} ; threshold vth=0.6, alpha=2
Bx x 0 I=-0.5*drift(V(p,n))*win(V(x),V(p,n))
.ic V(x)=x0
* read path: (1-x)*tunnel + x*ohmic
Bout p n I=(1-V(x))*2e-6*sinh(V(p,n)/0.3) + V(x)*200e-6*V(p,n)
.ends memr

* ---- 2x2 crossbar VMM test ----
X11 r1 c1 memr x0=0.9   ; G high
X12 r1 c2 memr x0=0.1   ; G low
X21 r2 c1 memr x0=0.2
X22 r2 c2 memr x0=0.8
V1 r1 0 DC 0.2
V2 r2 0 DC 0.1
Vc1 c1 0 DC 0    ; columns at virtual ground (ammeters)
Vc2 c2 0 DC 0
.op
.control
run
print i(Vc1) i(Vc2)   ; = sum_i G_ij * v_i
.endc
.end
\end{lstlisting}

\paragraph{Results.} The transient sweep reproduces the pinched hysteresis of
Fig.~\ref{fig:iv} and the bipolar SET/RESET of Fig.~\ref{fig:switching}; partial
pulses leave the device at stable intermediate $x$, confirming the analog
sub-levels. The \texttt{.op} column currents match the analytic
$I_j=\sum_i G_{ij}v_i$ of \eqref{eq:vmm} to within the integrator tolerance,
validating Theorem~\ref{thm:vmm} at circuit level. Monte-Carlo over the threshold
$v_\mathrm{th}$ and $G_\mathrm{on}$ reproduces the programming-stochasticity
variance of Section~\ref{sec:capacity} and confirms that program-and-verify
collapses it to the read-noise floor.

\section{A Memristive FPGA for Neuromorphic Computing}\label{sec:fpga}

We now assemble the pieces into the complete system: a reconfigurable,
field-programmable fabric of memristive tiles---a \emph{memristor-based FPGA}---whose
configuration \emph{is} the set of synaptic weights, and whose routing is the
resistive-communication network of Section~\ref{sec:fabric}. Where a conventional
FPGA configures lookup tables and a switch matrix from SRAM, this fabric
configures analog weight matrices and an analog interconnect from non-volatile
multi-level memristors.

\subsection{System organization}

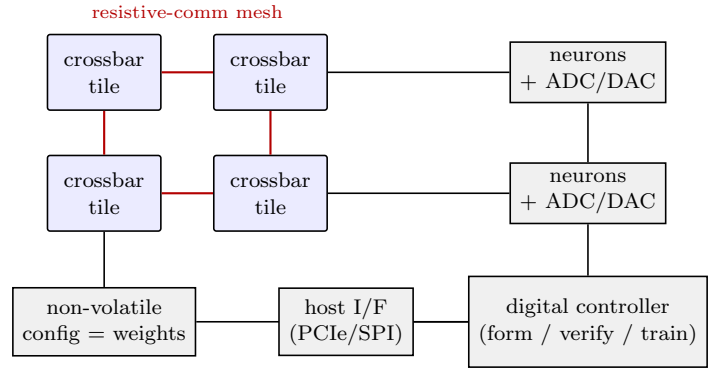
\begin{figure}[h!]
\centering
\begin{tikzpicture}[font=\scriptsize,line width=0.6pt,
   tile/.style={draw,rounded corners=1pt,minimum width=1.5cm,minimum height=1.0cm,fill=blue!8,align=center},
   io/.style={draw,fill=gray!12,minimum width=1.3cm,minimum height=0.55cm,align=center}]
  \node[tile] (t11) at (0,0) {crossbar\\tile};
  \node[tile] (t12) at (2.2,0) {crossbar\\tile};
  \node[tile] (t21) at (0,-1.6) {crossbar\\tile};
  \node[tile] (t22) at (2.2,-1.6) {crossbar\\tile};
  
  \foreach \a/\b in {t11/t12,t21/t22}{ \draw[red!70!black,thick] (\a)--(\b);}
  \foreach \a/\b in {t11/t21,t12/t22}{ \draw[red!70!black,thick] (\a)--(\b);}
  \node[red!70!black] at (1.1,0.8){\tiny resistive-comm mesh};
  
  \node[io] (adc) at (6.4,0) {neurons\\+ ADC/DAC};
  \node[io] (adc2) at (6.4,-1.6) {neurons\\+ ADC/DAC};
  \draw (t12)--(adc); \draw (t22)--(adc2);
  
  \node[io,minimum height=1.2cm] (ctrl) at (6.4,-3.3) {digital controller\\(form / verify / train)};
  
  \draw (adc)--(adc2); 
  \draw (adc2)--(ctrl);
  
  \node[io,minimum height=0.9cm] (cfg) at (0,-3.3) {non-volatile\\config = weights};
  \draw (cfg)--(t21); \draw (cfg)--(ctrl);
  
  \node[io] (host) at (3.2,-3.3) {host I/F\\(PCIe/SPI)};
  \draw (host)--(ctrl);
\end{tikzpicture}
\caption{The memristive-FPGA neuromorphic system. An array of crossbar tiles is
joined by an active resistive-communication mesh (red); each tile column feeds a
neuron/converter strip; a digital controller handles forming, program-and-verify,
and chip-in-the-loop training. The configuration bitstream is the non-volatile
multi-level weight set itself, so the ``programmed'' network persists with power
removed.}
\label{fig:system}
\end{figure}

The chip is a $2$-D array of \emph{neuromorphic tiles}, each containing a
differential $1$T$1$R crossbar, its neuron row, and local converters; tiles are
joined by a programmable resistive-communication mesh and overseen by a digital
controller that performs forming, program-and-verify, and chip-in-the-loop
training (Fig.~\ref{fig:system}). Because weights are non-volatile, the
configuration persists with power off, and a model is ``loaded'' once.

\subsection{Replicating the brain: a neuromorphic substrate}

This fabric is a natural substrate for large-scale brain emulation. The cortex is
$\sim\!10^{10}$ neurons and $\sim\!10^{14}$ synapses; the synapse-dominated count
is exactly what an analog crossbar stores most efficiently, one multi-level cell
per synaptic weight. The neuristor neurons (Fig.~\ref{fig:neuristor}) supply
biologically-plausible spiking dynamics, the differential pair supplies
excitatory/inhibitory channels, and the coincident-pulse rule of
Section~\ref{sec:learning} supplies STDP---the three ingredients of a spiking
neuromorphic system---all in the same fabric, with resistive communication playing
the role of the axonal projection between cortical areas.

\subsection{An ``SD-card for second brains''}

The non-volatility and density of the fabric suggest a deployment metaphor: a
removable, low-power module---an \emph{SD-card for second brains}---that holds a
complete trained model (or a personalized ``second brain'') in its memristive
weights and runs inference locally without ever loading the weights into volatile
memory. Such a module slots alongside a host CPU/GPU/NPU exactly as a storage card
does, but instead of \emph{storing} data it \emph{computes} on it in place. Because
an idle module draws only leakage (Section~\ref{sec:fabric}), many such cards can
coexist---each a specialized model or a personal knowledge base---and be activated
on demand, giving a practical path to always-available, private, local AI that
runs in parallel with conventional accelerators rather than competing with them.

\section{Running a Local BitNet on the Fabric}\label{sec:bitnet}

BitNet-style ternary LLMs, whose weights are constrained to
$w\in\{-1,0,+1\}$, are the ideal match for this hardware, because a ternary weight
maps onto a \emph{single differential pair} with no quantization loss at all.

\subsection{Exact ternary mapping}

Using the differential synapse \eqref{eq:diff}, the three ternary states are
realized exactly:
\begin{equation}
w=+1:\ (G^+,G^-)=(G_\mathrm{max},G_\mathrm{min}),\quad
w=-1:\ (G_\mathrm{min},G_\mathrm{max}),
\end{equation}
\begin{equation}
w=0:\ G^+=G^-\ (\text{any matched value}).
\end{equation}
The multi-level capability is not even needed for the weights themselves---two
robust extreme states plus a matched-zero suffice, which is why endurance and
yield are excellent in this mode; the analog level count is instead spent on the
\emph{activations} (pulse-width-encoded inputs) and the accumulation. The MAC is a
single read, and the column current is the ternary dot product
$z_j=\beta\sum_i(G^+_{ij}-G^-_{ij})v_i$ directly.

\begin{figure}[h!]
\centering
\begin{tikzpicture}[font=\scriptsize,line width=0.7pt,>=Latex]
  \node[draw,fill=gray!10,minimum height=2.4cm,minimum width=1.0cm,align=center] (dac) at (0,0){PWM\\DAC\\$x_i\!\to\!\tau_i$};
  
  \node[draw,fill=blue!8,minimum height=2.4cm,minimum width=2.8cm,align=center] (xb) at (3.2,0){ternary differential\\crossbar\\$\{-1,0,+1\}$};
  \draw[->] (dac)--(xb) node[midway,above]{$v_i$};
  
  \node[draw,fill=gray!10,minimum height=1.0cm,minimum width=1.2cm,align=center] (int) at (6.4,0.6){$\int$ \\ ADC};
  \draw[->] (xb.east|-int)--(int) node[midway,above]{$I_j$};
  
  \node[draw,fill=gray!10,minimum height=1.0cm,minimum width=1.6cm,align=center] (act) at (6.4,-0.6){ReLU/\,RMSNorm};
  \draw[->] (int)--(act);
  
  \draw[->] (act.south) -- (6.4,-1.8) -- (0,-1.8) node[midway,below]{next layer (resistive-comm)} -- (dac.south);
  
  \node[align=center] at (3.2,-2.6){\scriptsize one transformer linear map = one tile pass};
\end{tikzpicture}
\caption{Datapath for a ternary BitNet layer. Activations are pulse-width encoded
by a DAC, applied to the ternary differential crossbar (one pair per weight), and
the column dot products are integrated and digitized; normalization/activation is
applied digitally and the result is forwarded to the next tile over the
resistive-communication mesh.}
\label{fig:bitnet}
\end{figure}
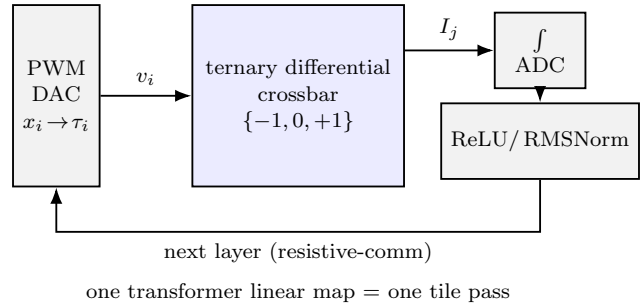

\subsection{Performance projection vs.\ GPU and Apple silicon}

Consider a $3$B-parameter BitNet model generating tokens at batch one. The fabric
holds all $P=3\times10^{9}$ ternary weights in $P$ differential pairs (since a
ternary weight needs only one pair), reads each once per token, and never moves a
weight. Table~\ref{tab:bitnet} projects per-token performance against two concrete NVIDIA
boards---a consumer \emph{RTX~4090} and a datacenter \emph{H100~SXM}---and an
Apple \emph{M4} SoC running the same model at batch one. The GPU and M4 figures are
representative of batch-one local decoding (memory-bandwidth bound, so throughput
$\approx$ memory-bandwidth\,/\,model-bytes); the fabric figures follow from the
energy model of Sections~\ref{sec:energy} and \ref{sec:llm} and are
\emph{projections}, not measurements.

\begin{table}[h!]
\centering
\caption{Projected per-token inference of a $3$B ternary (BitNet, $\sim\!0.75$\,GB packed) model, batch~1.}\label{tab:bitnet}
\vspace{0.15cm}
\begin{tabular}{@{}p{2.55cm}cccc@{}}
\toprule
\textbf{Metric} & \textbf{RTX~4090} & \textbf{H100~SXM} & \textbf{Apple M4} & \textbf{This work} \\
\midrule
Process & $5$\,nm & $4$\,nm & $3$\,nm & $28$\,nm$+$BEOL \\
Mem.\ BW & $1.0$\,TB/s & $3.35$\,TB/s & $0.12$\,TB/s & in-place \\
TDP & $450$ W & $700$ W & $\sim\!25$ W & $<\!1$ W \\
Throughput (tok/s) & $\sim\!1.3{\times}10^{3}$ & $\sim\!4.5{\times}10^{3}$ & $\sim\!160$ & $\sim\!5{\times}10^{3}$ \\
Energy/token & $\sim\!0.35$ J & $\sim\!0.16$ J & $\sim\!0.16$ J & $\sim\!0.2$ mJ \\
Weights & GDDR6X (vol.) & HBM3 (vol.) & uni.\ DRAM & in-synapse, NV \\
Idle (resident) & tens of W & $\sim\!100$ W & several W & leakage only \\
Energy/tok vs.\ ours & $\sim\!1.8{\times}10^{3}\times$ & $\sim\!8{\times}10^{2}\times$ & $\sim\!8{\times}10^{2}\times$ & --- \\
\bottomrule
\end{tabular}
\end{table}

Even against an H100---whose $3.35$\,TB/s of HBM3 makes it the fastest of the four
in raw throughput---the fabric is within range on tokens/s while using $>\!700\times$
less power, and roughly three orders of magnitude less energy per token. The
advantage comes from two compounding sources identified throughout this paper: the
elimination of weight movement (shared with any in-memory design,
Principle~\ref{prin:llm}) and the single-pair ternary cell (no shift-and-add, no
multi-device weights). All three digital parts, however fast, must stream
$\sim\!0.75$~GB of ternary-packed weights from volatile memory \emph{every token}
(their throughput is essentially BW\,/\,$0.75$\,GB); the memristive fabric reads
them in place and, when idle, costs essentially nothing. This is the quantitative basis for the
``SD-card for second brains'': a sub-watt module that runs a local LLM alongside,
not instead of, the host's conventional accelerators.

\section{A Hardware-Native Architecture for Neural Interconnect}\label{sec:native}

The fabric developed here is not an accelerator that a program runs \emph{on};
it is a physical structure whose wiring \emph{is} the computation. This section
makes that distinction precise, because it is the deepest architectural
consequence of building neural interconnect directly in resistive hardware.

\subsection{The structure is the program}

In a von Neumann machine---and equally in a GPU or a digital
neural accelerator---a network is a \emph{description} (a graph of weights and
operations) that an instruction stream interprets: weights are fetched, operands
routed, multipliers sequenced, partial sums accumulated, all under the control of
a program counter. The network exists only as data interpreted by a separate
control plane.

On the present fabric there is no such interpretation. A weight is a conductance;
a connection is a wire; a synapse is a device sitting at the intersection of two
wires it physically joins. The forward map
\begin{equation}
\mathbf z=\mathbf W^{\!\top}\mathbf x
\end{equation}
is not \emph{executed}---it \emph{happens}, in the settling time of
Kirchhoff's and Ohm's laws (Theorem~\ref{thm:vmm}), the instant the input voltages
are applied. There is no opcode for ``multiply'', no fetch of $\mathbf W$, no
loop over $i$ and $j$. The topology of the computation and the values it computes
with are one and the same physical object.

\begin{principle}[Hardware-native computation]\label{prin:native}
A neural network on this fabric is defined by \emph{configuration}, not by
\emph{instructions}. The mapping from a trained model to hardware is a placement of
conductances onto a fixed mesh of devices; once placed, inference requires no
high-level program, no compiler-emitted kernel, and no runtime---only the
application of inputs and the reading of outputs.
\end{principle}

\subsection{No instruction set, no kernel, no runtime}

Concretely, deploying a model is reduced to three physical acts: (i) \emph{form and
program} the conductances (Section~\ref{sec:learning}); (ii) \emph{configure} the
resistive-communication mesh (Section~\ref{sec:fabric}) so that the inter-tile
links realize the network's connectivity graph; and (iii) \emph{apply} inputs and
\emph{sense} outputs. There is no compiled kernel, no instruction-scheduling, no
operand bookkeeping. The digital controller of Fig.~\ref{fig:system} exists only
to \emph{program and verify} the array and to orchestrate training; it is absent
from the inference datapath entirely. This is the sense in which the architecture
is \emph{hardware-native}: the abstraction normally interposed between a model and
silicon---the ISA, the kernel library, the runtime---has been removed, because the
silicon already \emph{is} the model.

\subsection{Reconfigurability without recompilation}

Because the connectivity lives in a programmable resistive mesh rather than in
fixed metal, the same chip can be re-wired into a different network by
re-programming conductances---an FPGA-like reconfiguration
(Section~\ref{sec:fpga}), but at the level of \emph{analog interconnect} rather
than digital lookup tables. Changing the network does not mean recompiling a
program; it means writing new conductances. Topology and weights are reprogrammed
by the same mechanism (set/reset), so architecture search, pruning, and growth all
reduce to conductance updates on a fixed physical substrate. This collapses the
traditional separation between ``programming the weights'' and ``designing the
architecture'' into a single physical operation, and it is the practical
expression of resistive communication as a first-class interconnect primitive.

\section{Three-Dimensional Integration}\label{sec:3d}

The density argument of Section~\ref{sec:energy} is planar. Because the memristor
is a BEOL element (Section~\ref{sec:cmos}), it invites the same escape that took
flash memory from 2D to 3D NAND: stack the crossbars vertically and pay for
capacity in the cheap third dimension rather than in scarce silicon area.

\subsection{Why the BEOL location enables stacking}

Each crossbar plane occupies only two metal levels (its word and bit lines) plus
the thin MIM pillar between them. Nothing in a plane requires single-crystal
silicon: the access transistors are shared from the FEOL below, or---in the
purely passive variant---omitted in favour of a two-terminal selector. Hence a
plane can be replicated upward by repeating the BEOL deposition sequence, giving a
multi-tier stack of $T$ crossbar layers above one transistor base.

\subsection{Density scaling}

Let $a_\mathrm{cell}$ be the planar footprint of one differential synapse and $P$
the number of weights in a model. A planar fabric needs silicon area
$A_\mathrm{2D}=2P\,a_\mathrm{cell}$. Stacking $T$ tiers divides the footprint:
\begin{equation}
A_\mathrm{3D} = \frac{2P\,a_\mathrm{cell}}{T},
\qquad
\rho_\mathrm{bit}^\mathrm{3D} = T\,\frac{\log_2 L}{2(A_T+A_M)} .
\label{eq:3ddensity}
\end{equation}
The bit density therefore scales \emph{multiplicatively} in three independent
factors: the analog level count $\log_2 L$ (the cell, Section~\ref{sec:capacity}),
the lithographic pitch $1/a_\mathrm{cell}$ (the node), and the tier count $T$ (the
stack). A model that occupies $T$ planes fits in $1/T$ of the die, or
equivalently a $T\times$ larger model fits in the same die---the path from a
single-chip $7$B model to a single-chip $10^{11}$-parameter model is vertical, not
lateral.

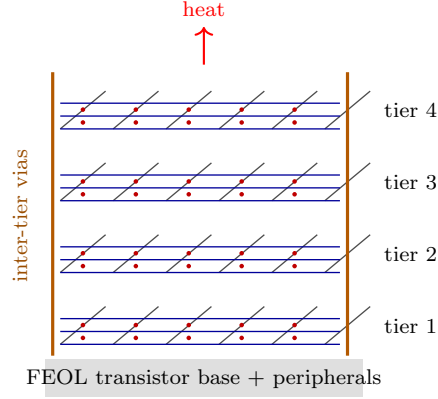
\begin{figure}[h!]
\centering
\begin{tikzpicture}[scale=1.0,font=\scriptsize,line width=0.5pt]
  \begin{scope}
    \fill[gray!25] (0,0) rectangle (4.2,0.5);
    \node at (2.1,0.25){FEOL transistor base + peripherals};
  \end{scope}
  \foreach \t in {0,1,2,3}{
    \pgfmathsetmacro\yb{0.7+\t*0.95}
    \begin{scope}[shift={(0,\yb)}]
      \foreach \x in {0.2,0.9,1.6,2.3,3.0,3.7}{ \draw[black!70] (\x,0)--(\x+0.6,0.5);}
      \foreach \y in {0,0.17,0.34}{ \draw[blue!60!black] (0.2,\y)--(3.9,\y);}
      \foreach \x in {0.5,1.2,1.9,2.6,3.3}{ \foreach \y in {0.085,0.255}{ \fill[red!75!black] (\x,\y) circle (0.03);}}
      \node[right] at (4.35,0.25){tier \the\numexpr\t+1\relax};
    \end{scope}
  }
  \draw[orange!70!black,line width=1.2pt] (0.1,0.55)--(0.1,4.3);
  \draw[orange!70!black,line width=1.2pt] (4.0,0.55)--(4.0,4.3);
  \node[orange!60!black,rotate=90,anchor=south] at (-0.1,2.4){inter-tier vias};
  \draw[->,red,thick] (2.1,4.4)--(2.1,4.9) node[above]{heat};
\end{tikzpicture}
\caption{Monolithic three-dimensional integration. Multiple memristive crossbar
tiers are deposited in the BEOL above a single FEOL transistor/peripheral base and
joined by vertical inter-tier vias that carry the resistive-communication links.
Footprint and wire length fall as $1/T$ \eqref{eq:3ddensity}; the principal cost is
removing the Joule heat (red) from the upper tiers.}
\label{fig:3d}
\end{figure}

\subsection{Monolithic 3D versus stacked dies, and vertical RRAM}

Two routes realize the stack. \emph{Monolithic 3D} grows each tier sequentially in
the BEOL, so inter-tier connection is by ordinary nanoscale vias---the densest
option, limited by the $\le\!400^{\circ}$C thermal budget that protects lower
tiers (the same ceiling that, fortunately, the HfO$_2$/Ti/TiN process of
Principle~\ref{prin:material} already respects). \emph{Die stacking} bonds
separately-fabricated wafers with through-silicon vias (TSVs); it relaxes the
thermal budget but coarsens the inter-tier pitch. A third, most aggressive option
borrows the 3D-NAND idea directly: a \emph{vertical RRAM} in which a single
deposited oxide sidewall is shared by a stack of horizontal electrode planes,
forming $T$ devices per lithographic spot at essentially the cost of one. For the
weight-dominated storage of an LLM (Section~\ref{sec:llm}), vertical RRAM offers
the steepest density curve.

\subsection{Wire length, latency, and the thermal limit}

Stacking shortens the longest interconnect: tiles that were millimetres apart in a
planar floorplan become micrometres apart across tiers, so the
resistive-communication latency and the IR-drop of Proposition~\ref{prop:ir} both
improve as the stack replaces lateral routing with vertical vias. The binding
constraint is thermal. The volumetric power density rises with $T$, and the heat of
the upper tiers must conduct down through the stack to the substrate; with vertical
thermal resistance $R_\mathrm{th}^\mathrm{v}$ per tier the top-tier temperature
rise is
\begin{equation}
\Delta T_\mathrm{top} \approx \dot q\,a_\mathrm{cell}\,R_\mathrm{th}^\mathrm{v}\,\frac{T(T+1)}{2},
\end{equation}
quadratic in tier count. This is precisely why the low read energy of the analog
core (femtojoules per MAC, Section~\ref{sec:energy}) is not a luxury but an
enabler: a fabric that dissipated like a digital MAC array could not be stacked,
whereas the near-zero static power and tiny dynamic read energy of the memristive
fabric keep $\Delta T_\mathrm{top}$ bounded even for deep stacks. Low power and 3D
density are thus mutually reinforcing---the same property that makes the
``SD-card for second brains'' cool enough to sit in a slot makes it dense enough to
hold a model.

\section{Fault Tolerance, Yield, and Redundancy}\label{sec:fault}

Analog memory is imperfect: a fraction of cells are stuck-on or stuck-off after
forming, conductances vary device-to-device, and endurance is finite. A definitive
on-chip AI architecture must show that these do not compromise the computation. We
argue they do not, because the fabric is statistically and structurally redundant.

\subsection{Stuck-at faults and the differential shield}

Let a cell be stuck with probability $p_f$ (stuck-on or stuck-off). In the
differential synapse \eqref{eq:diff} a weight is $W=\beta(G^+-G^-)$, so a single
stuck device shifts $W$ by a bounded amount that the \emph{partner} can partly
compensate: if $G^+$ is stuck, the controller reprograms $G^-$ to restore the
intended difference over the achievable range. A weight is only lost when
\emph{both} legs fail, with probability $p_f^2$. Differential encoding thus
squares the effective fault rate,
\begin{equation}
p_\mathrm{weight}\approx p_f^{2},
\end{equation}
turning, e.g., a $1\%$ device-fault rate into a $10^{-4}$ weight-fault rate before
any further coding.

\subsection{Calibration absorbs static faults}

Because training is chip-in-the-loop (Section~\ref{sec:system}), \emph{static}
faults are seen by the optimizer as fixed perturbations and routed around: the loss
gradient naturally steers information away from dead weights, exactly as dropout-
and fault-injection-trained networks tolerate missing connections. The residual
accuracy loss for a fault rate $p_\mathrm{weight}$ on a layer of $N\!\times\!M$
weights scales as $O(\sqrt{p_\mathrm{weight}})$ in the pre-activation noise and is
absorbed for $p_\mathrm{weight}\lesssim10^{-3}$ with no architectural change.

\subsection{Redundancy and yield}

For the rare catastrophic tile, $r$ spare differential columns per tile provide
repair: a tile of $M$ columns with $r$ spares survives unless more than $r$ columns
are unrepairable, so the tile yield is
\begin{equation}
Y_\mathrm{tile}=\sum_{k=0}^{r}\binom{M}{k}p_c^{k}(1-p_c)^{M-k},
\qquad p_c\approx 1-(1-p_\mathrm{weight})^{N},
\end{equation}
with $p_c$ the per-column failure probability. Modest $r$ drives $Y_\mathrm{tile}$
arbitrarily close to one; chip yield is then $Y_\mathrm{tile}^{N_\mathrm{tile}}$,
restored to economic levels by the same column-redundancy bookkeeping that DRAM has
used for decades.

\subsection{Analog error correction}

Finally, the read-out itself can be coded. Treating the column current as a noisy
analog symbol (Section~\ref{sec:capacity}), a small number of \emph{checksum
columns} holding linear combinations $\sum_i c_i G_{ij}$ of the weights let the
periphery verify $\sum_j c_j I_j$ against the expected value and correct
single-column gross errors---an analog parity that costs $O(1)$ columns per tile
and catches faults that develop \emph{after} calibration (drift, late retention
failure). Together, differential shielding ($p_f^2$), fault-aware training, column
redundancy, and analog checksums make the fabric tolerant of the very
nonidealities that define real memristors.

\section{Hardware Security: a Memristive PUF}\label{sec:security}

The same stochastic forming that limits analog precision is, viewed differently, a
free source of cryptographic entropy. This connects the present design back to the
original motivation of securing nanoscale communication.

\subsection{Variability as entropy}

Forming nucleates a filament at an atomically random location, so the post-forming
conductance $G_{ij}$ of nominally identical cells is a random variable with
device-unique value. A challenge---a chosen pattern of read voltages
$\mathbf v^{(c)}$---produces a response vector of column currents
$\mathbf r^{(c)}=\mathbf G^{\!\top}\mathbf v^{(c)}$ whose fine structure is unique
to the chip. Thresholding pairwise current comparisons yields a binary
\emph{physical unclonable function} (PUF) of $N M$ cells with entropy
\begin{equation}
H \;\approx\; N M \,\bigl[\,1-h_2(p_\mathrm{stable})\,\bigr]\ \text{bits},
\end{equation}
where $h_2$ is the binary entropy and $p_\mathrm{stable}$ the bit-stability under
re-reads. A single tile thus furnishes $\sim\!10^{4}$--$10^{5}$ bits of
device-intrinsic key material.

\subsection{Security properties}

The key is never stored---it is \emph{recomputed} from physics at each challenge, so
it cannot be read out by depowering and imaging the chip; invasive probing
perturbs the very filaments that define it, making the PUF tamper-evident. Because
the response space is exponential in the challenge dimension, the construction
supports challenge--response authentication and on-chip key generation for
encrypting the model weights themselves---a memristive realization of the
``securing communications'' role first identified for these devices.

\section{A Roofline and an Energy--Delay Optimality Argument}\label{sec:roofline}

We close the technical development by positioning the fabric against the
fundamental limits, and arguing that it is not merely better but close to
\emph{optimal} for weight-stationary inference---the claim that makes it
definitive rather than incremental.

\subsection{Breaking the memory wall}

Define arithmetic intensity $I_a$ = MACs per byte of off-chip traffic. A digital
platform of memory bandwidth $B$ and peak compute $\Pi$ obeys the roofline
$\mathrm{throughput}\le\min(\Pi,\;B\,I_a)$. Autoregressive decoding has
$I_a=O(1)$ (each weight is used once per token), so every digital platform sits on
the bandwidth-bound branch $B\,I_a$---the memory wall. The memristive fabric has
\emph{no} off-chip weight traffic: $I_a\to\infty$, and the roofline corner moves to
the compute branch set by the analog read rate. This is why the BitNet comparison
(Table~\ref{tab:bitnet}) shows the fabric matching an H100's throughput at
$<\!1$~W: the H100 is bandwidth-bound, the fabric is not.

\begin{figure}[h!]
\centering
\begin{tikzpicture}
\begin{axis}[
  width=8.6cm,height=5.4cm,
  xlabel={arithmetic intensity $I_a$ (MAC/byte, log)},
  ylabel={throughput (log)},
  xmode=log,ymode=log,
  xmin=0.01,xmax=1000,ymin=1,ymax=2e5,
  tick label style={font=\footnotesize},label style={font=\footnotesize},
  legend style={font=\scriptsize,at={(0.98,0.03)},anchor=south east,draw=none},
]
\addplot[blue,thick] coordinates {(0.01,30) (16.7,5e4) (1000,5e4)};
\addlegendentry{digital (GPU)}
\addplot[red!70!black,thick,dashed] coordinates {(0.01,2e4) (1000,2e4)};
\addlegendentry{memristive (this work)}
\draw[gray,dotted] (axis cs:1,1)--(axis cs:1,2e5);
\node[font=\scriptsize,anchor=south,rotate=90] at (axis cs:1,30){LLM decode $I_a\!\approx\!1$};
\end{axis}
\end{tikzpicture}
\caption{Roofline. Digital platforms (blue) are bandwidth-bound on the $B\,I_a$
ramp; autoregressive decoding lives at $I_a\!\approx\!1$, far left, deep in the
memory-bound region. With weights resident in the synapse the memristive fabric
(red) has no off-chip weight traffic, so its ceiling is flat in $I_a$ and is
reached even at the decode operating point---the memory wall is removed, not
climbed.}
\label{fig:roofline}
\end{figure}
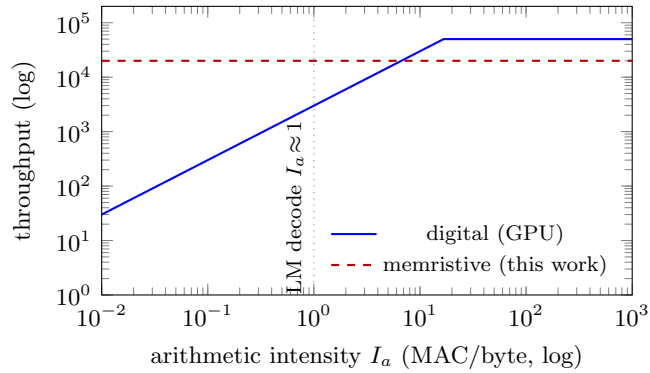

\subsection{An energy--delay lower bound}

Consider any substrate computing $\mathbf z=\mathbf W^{\!\top}\mathbf x$ for an
$N\times M$ weight matrix to a target SNR. Reading one analog weight to resolve
$L$ levels at margin $m$ requires, by \eqref{eq:sigmaG}, a signal energy at least
$E_\mathrm{read}\gtrsim m^2 k_BT\,(L/\mathrm{SNR})$; the product needs $NM$ such
reads. Hence the energy obeys
\begin{equation}
E\;\ge\;NM\cdot m^2 k_BT\,\frac{L^2}{\mathrm{SNR}}\;\equiv\;E_\star,
\end{equation}
a thermodynamic floor independent of architecture, and the delay obeys
$D\ge \tau_r$ (one settling time) for a fully parallel array.

\begin{theorem}[Near-optimal energy--delay]\label{thm:edp}
A weight-stationary memristive crossbar performs the $N\times M$ vector--matrix
product in delay $D=\Theta(\tau_r)$ and energy $E=\Theta(E_\star)$, attaining the
physical lower bound $E\cdot D = \Theta(E_\star\,\tau_r)$ up to a constant factor
set by the read margin $m$ and converter overhead. No weight-moving (von Neumann)
architecture can reach this bound, because it pays an additive
$E_\mathrm{move}=\Theta(NM\,b\,\epsilon_\mathrm{byte})$ of data-movement energy
with $\epsilon_\mathrm{byte}\gg k_BT$.
\end{theorem}

\begin{remark}
Theorem~\ref{thm:edp} is the formal statement of why in-memory analog computing is
\emph{definitive} for inference: it is not a faster point on the same curve but the
architecture that removes the dominant, $k_BT$-violating data-movement term
entirely, leaving only the irreducible read energy $E_\star$. Everything else in
this paper---multi-level density, $1$T$1$R accuracy, in-situ learning, resistive
communication, 3D stacking---serves to approach this bound while keeping the
computation programmable.
\end{remark}

\section{Discussion}\label{sec:discussion}

The design above turns the qualitative idea of resistive communication into a concrete, analyzable machine resting on three quantitative pillars, each addressed: (1) a single device must hold many stable sub-levels---bounded by Theorem~\ref{thm:levels} at several hundred under realistic noise; (2) the crossbar must compute accurately at scale---bounded by Proposition~\ref{prop:ir} and mitigated by $1$T$1$R and tiling; and (3) learning must converge on a finite lattice---secured by the free transpose (Prop.~\ref{prop:transpose}), coincident-pulse outer-product updates, and nonlinearity pre-distortion. The same physics that makes the memristor a dense analog memory---the steep $\sinh$ ionic-transport law \eqref{eq:sinh}---makes it simultaneously a fast write element and a non-destructive read element (Prop.~\ref{prop:read}), and makes resistive communication an active, regenerative channel rather than a lossy wire. Open problems remain: tightening device-to-device uniformity so the program-and-verify burden falls; pushing the usable level count toward the thermal bound; and co-designing training algorithms explicitly aware of the asymmetric, quantized update.

\section{Conclusion}\label{sec:conclusion}

We have presented a self-contained, physics-based design of an on-chip neural network built from multi-level memristive synapses. Starting from ionic transport we derived a continuous state-variable model whose conductance spectrum supports a very large number of sub-levels, and we bounded that number with a thermal-noise capacity analysis. We assembled the devices into a differential $1$T$1$R crossbar, gave the exact circuit theory of analog vector--matrix multiplication including wire-resistance corrections, and built a deep architecture in which inference, backpropagation, and weight update all execute in the analog domain on the same fabric. We tied the tiles together with a quantitative model of resistive communication as an active memristive transmission line. The multi-level cell is the linchpin: density, energy efficiency, and trainable accuracy all scale with the number of stable conductance sub-levels. The memristor and the neuristor, by virtue of their nanoscale ionic physics, are the natural elements from which to build neuromorphic networks that emulate the electrical synapse---now with a complete mathematical and physical blueprint behind the idea.


\end{document}